\documentclass[12pt]{article}
\usepackage{babel}
\usepackage{arxiv}
\usepackage{amssymb} 
\usepackage[utf8]{inputenc} 
\usepackage[T1]{fontenc}	
\usepackage{times}      	
\usepackage{url}        	
\usepackage{booktabs}   	
\usepackage{amsfonts}   	
\usepackage{nicefrac}   	
\usepackage[protrusion=true,expansion=true,verbose=false,disable=false,final,babel=true,tracking=false,patch=none]{microtype} 
\usepackage{graphicx} 	 
\usepackage{subcaption}
\usepackage{doi}
\usepackage{amsmath, calc}
\usepackage{float}
\usepackage{mathabx}
\usepackage{apacite}
\usepackage[toc,page]{appendix}
\usepackage{listings}
\usepackage{xcolor}
\usepackage{setspace}
\AtBeginDocument{\newgeometry{left=2cm,right=2cm,top=3cm,bottom=3cm}}
\usepackage{natbib}
\usepackage{multicol}
\usepackage{multirow}
\usepackage{array}
\usepackage{makecell}
\usepackage{colortbl}
\usepackage{longtable}
\usepackage{tikz}
\usepackage{arydshln}
\usepackage{siunitx}
\usepackage{soul}
\usepackage[most]{tcolorbox}
\definecolor{lightgreen}{RGB}{204,255,204}  
\definecolor{lightyellow}{RGB}{255,255,204}
\sethlcolor{lightyellow}
\usetikzlibrary{arrows, positioning} 

\definecolor{codegreen}{rgb}{0,0.6,0}
\definecolor{codegray}{rgb}{0.5,0.5,0.5}
\definecolor{codepurple}{rgb}{0.58,0,0.82}
\definecolor{backcolour}{rgb}{0.95,0.95,0.92}

\lstdefinestyle{mystyle}{
	backgroundcolor=\color{backcolour},   
	commentstyle=\color{codegreen},
	keywordstyle=\color{magenta},
	numberstyle=\tiny\color{codegray},
	stringstyle=\color{codepurple},
	basicstyle=\ttfamily\footnotesize,
	breakatwhitespace=false,   	 
	breaklines=true,           	 
	captionpos=b,              	 
	keepspaces=true,           	 
	numbers=left,              	 
	numbersep=5pt,            	 
	showspaces=false,          	 
	showstringspaces=false,
	showtabs=false,            	 
	tabsize=2
}

\title{Characterising mortality dynamics across countries and time using a multi-stage clustering approach}

\date{}

\author{ Pedro Menezes de Araújo\\
	School of Mathematics and Statistics\\
	University College Dublin\\
	\texttt{pedro.menezesdearaujo@ucdconnect.ie} \\
	\And
	Ugofilippo Basellini \\
	Max Planck Institute for Demographic Research\\
	\texttt{basellini@demogr.mpg.de} \\
	\And
	Thomas Brendan Murphy \\
	School of Mathematics and Statistics\\
	University College Dublin\\
	\texttt{brendan.murphy@ucd.ie} \\
	\And
	Isobel Claire Gormley \\
	School of Mathematics and Statistics\\
	University College Dublin\\
	\texttt{claire.gormley@ucd.ie} \\
}

\renewcommand{\headeright}{}
\renewcommand{\undertitle}{}

\makeatletter
\renewcommand{\@maketitle}{%
	\vbox{%
		\hsize\textwidth
		\linewidth\hsize
		\vskip 0.1in
		\centering
		{\LARGE\bfseries \@title\par}
		\textsc{\undertitle}\\
		\vskip 0.1in
		\def\And{%
			\end{tabular}\hfil\linebreak[0]\hfil%
			\begin{tabular}[t]{c}\bf\rule{\z@}{24\p@}\ignorespaces%
		}
		\def\AND{%
			\end{tabular}\hfil\linebreak[4]\hfil%
			\begin{tabular}[t]{c}\bf\rule{\z@}{24\p@}\ignorespaces%
		}
		\begin{tabular}[t]{c}\bf\rule{\z@}{24\p@}\@author\end{tabular}%
		\vskip 0.4in \@minus 0.1in \center{\@date} \vskip 0.2in
	}
}
\makeatother

\hypersetup{
pdftitle={Characterising mortality dynamics using a multi-stage clustering approach},
pdfkeywords={Beta Latent Variable Model, Hamiltonian Monte Carlo,  Human Mortality Database},
}
\DisableLigatures{encoding = *, family = *}

\begin{document}

\maketitle
\enlargethispage{3\baselineskip}

\begin{abstract} 
	Comparative analyses of mortality dynamics across countries have long shaped our understanding of mortality inequalities and patterns of divergence and convergence over time. However, most existing studies focus on either mortality differences across countries at a single point in time or on country trajectories, without considering both factors simultaneously. In this paper, we address this gap by introducing a multi-stage clustering framework, applied to 94 countries by sex over the period 1960-2019 using data from the World Population Prospects. Specifically, we model life-table probabilities of death using a time-dependent beta latent variable model and identify a small set of ``mortality states'' that characterise the mortality pattern of each country at any given point in time. We then cluster countries based on their sequences of mortality states. Our results reveal substantial heterogeneity in the timing and pace of transitions between mortality states, including a persistent East-West divide in Europe and distinctive Latin American patterns associated with elevated young-adult male mortality. Cross-country inequality increased for both sexes until the 1990s, before plateauing and subsequently declining. Our proposed multi-stage clustering framework provides an interpretable, time-explicit description of mortality dynamics that jointly captures within-country change and between-country differences, and can be applied to other settings where trajectories of categorical or discretised profiles are of interest. 
\end{abstract}


\newpage


\newpage

\section{Introduction}

The study of mortality patterns over age, time, and populations has long been a central topic in demography, epidemiology, and actuarial science. Although several regularities in mortality age patterns and time trends have been documented \citep{gompertz_nature_1825,tuljapurkar_universal_2000}, substantial heterogeneity in mortality has characterised populations historically and persists today \citep{riley_rising_2001, mesle_historical_2011}. Identifying groups of countries that exhibit similar mortality profiles and temporal dynamics is therefore valuable for a range of applications, including indirect mortality estimation, the study of mortality inequality and mortality forecasting. Historically, such assessments relied largely on visual inspection of mortality patterns. More recently, statistical methods have provided a more systematic foundation for studying mortality dynamics, with cluster analysis emerging as a key tool.

In recent years, many studies have applied clustering techniques to cross-national mortality data (for a comprehensive review, see \cite{Araujo_2025_review}). These studies aim to understand how mortality profiles differ across countries or how such differences evolve over time by analysing temporal clustering patterns, offering insight into both mortality transitions and emerging inequalities. Many contributions cluster mortality patterns across countries over time, either independently at a few time points \citep{Mesle_Vallin_2002, Atance_2024} or across larger periods of time \citep{Leger_Mazzuco_2021}. Alternatively, some studies cluster only the countries themselves \citep{Debon_Chaves_2017, Dimai_2025}, with additional analyses required to describe temporal changes in mortality and to characterise the resulting clusters. 

Taken together, existing research has focused either on clustering countries according to their mortality patterns at distinct time points or on clustering countries, with both approaches requiring additional analysis to fully understand relationships between countries or how mortality profiles changed over time. While both approaches have their own merits and purposes, neither fully captures the dual nature of mortality dynamics: the changes in mortality patterns within countries over time and the differences between countries. 

To address this gap, we propose a multi-stage clustering framework that simultaneously captures both aspects. First, we identify discrete ``mortality states'' by clustering country-year observations, effectively summarising the large heterogeneity of mortality curves into a small number of cluster-specific mortality patterns. The mortality states are derived by applying k-means clustering \citep{Hartigan_1979} to a lower-dimensional representation of mortality curves, obtained from a time-dependent beta latent variable model \citep{Araujo_2026_blv}. Second, we cluster countries based on their sequences of mortality states over time, allowing us to characterise groups of countries with similar mortality trajectories. This is achieved through a novel mixture of categorical distributions with group-specific penalised B-spline terms \citep{Hastie_1990_GAM}.
This approach captures smooth temporal evolution in the probability of occupying each mortality state, thereby enabling us to identify when major transitions occur within country clusters. Overall, this two-stage approach provides a comprehensive and interpretable description of mortality evolution, inequality, and country-level similarities over time. We apply our approach to 94 countries by sex using data from the 2024 revision of the World Population Prospects (WPP) dataset \citep{UN_2024} to provide a broad global perspective on mortality dynamics.


The remainder of this paper is organised as follows. Section~\ref{sec:wpp-data} describes the WPP data used in this study. Section~\ref{sec:dim-red} describes the dimensionality reduction procedure and Section~\ref{sec:clustering} presents our clustering methodology. In Section~\ref{sec:application}, we apply the multi-stage clustering method to the WPP dataset and discuss the results. Finally, Section~\ref{sec:conclusion} concludes with a discussion of limitations and future research directions. The \texttt{R} code used for all analyses is available in
\href{https://github.com/pedroaraujo9/clustering-mortality-two-step}{\url{https://github.com/pedroaraujo9/clustering-mortality-two-step}}.

\section{World Population Prospects life tables}\label{sec:wpp-data}

We employ period life table probabilities of death from the 2024 revision of the World Population Prospects (WPP) dataset \citep{UN_2024}, produced by the United Nations Population Division (UNPD). Let $q_{xit}$ denote the probability of death for age group $x$, country $i$, at time $t$. We use age groups starting from $[0, 1)$, $[1, 5)$, and $[5, 10)$ through $[80, 85)$, resulting in 18 age groups. We exclude older age groups due to concerns about the quality of mortality probability estimates at advanced ages, especially in low- and middle-income countries. 

The WPP provides life tables from 1950 onwards, and we use yearly data from 1960 to 2019. This time window allows us to avoid the COVID-19 pandemic, which we do not intend to investigate, while also avoiding the earliest years of the series, when data quality is uncertain, particularly for developing countries. The final dataset contains no missing observations and no probabilities equal to zero or one. 

Regarding countries, the UNPD classifies country data sources in the WPP as either ``Empirical'' or ``Model-based''. In the UNPD framework, ``Empirical'' sources are primarily based on vital registration data and may be supplemented by estimates and adjustments that are judged to be sufficiently reliable for reconstructing age- and sex-specific mortality. Those adjustments may include, for instance, old-age mortality adjustments, smoothing, interpolation for some gaps in the time series, and inclusion of crisis mortality impacts, with details available in~\cite{UN_2024}. On the other hand, ``Model-based'' countries have data that are too sparse or considered of very low quality, and most of the life tables are based on model life tables
\citep{Coale_Demeny_1966,coale_regional_1983}. We therefore include in our analysis only countries that are classified as ``Empirical'' to focus on countries with 
high- to mid-quality data. Additionally, we only consider countries with a population greater than 1 million (in 2023) to filter out mostly small island nations, which would require a weighting adjustment to account for the overrepresentation of mortality curves from such small populations. After applying 
both filters, the resulting dataset includes 94 countries. 

\subsection{Mortality trends}

\begin{figure}[H]
	\centering
	\includegraphics[width=\textwidth]{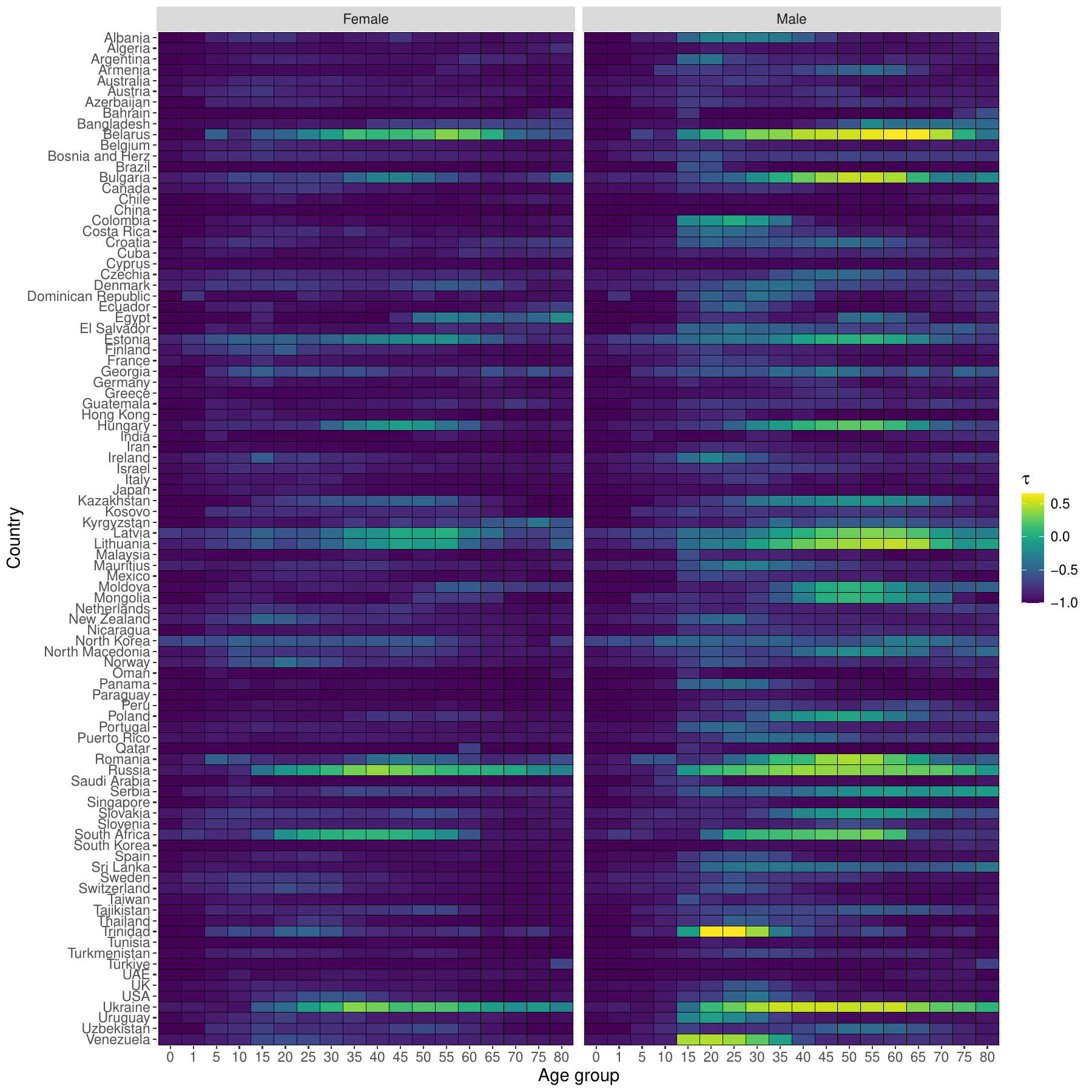}
	\caption{Kendall's $\tau$ trend coefficient for each country across age groups for females and males.}
	\label{fig:data:tau-wpp}
\end{figure}

All 94 countries used in our analysis are shown in Figure~\ref{fig:data:tau-wpp}, where we plot Kendall's $\tau$ trend coefficient \citep{Kendall_1976, Millard_2013} for each age group for females and males to summarise mortality information. The $\tau$ coefficient lies in the $[-1, 1]$ interval, being $-1$ for a perfect negative monotonic trend and $1$ for a perfect positive monotonic trend during the period 1960-2019. For females, the mortality trend is predominantly negative across all countries and age groups, with a few exceptions such as Belarus, Latvia, Lithuania, Russia, South Africa and Ukraine, which exhibit positive trends for some adult and old age groups. Some countries, such as Ireland, New Zealand, and Norway, show weaker negative trends, primarily because their mortality trajectories are more irregular due to smaller population sizes (around 5 million). For males, the diversity of mortality trends is considerably higher than for females. For young age groups (less than 15 years old), the trend is predominantly negative; however, for young adults, we observe countries with weaker negative trends, such as Brazil and Colombia, or even positive trends such as Trinidad and Venezuela. For older adults, Eastern European countries again display positive trends, with Belarus, Bulgaria, Lithuania, Romania, Russia, and Ukraine standing out.

\begin{figure}[H]
	\centering
	\includegraphics[width=\textwidth]{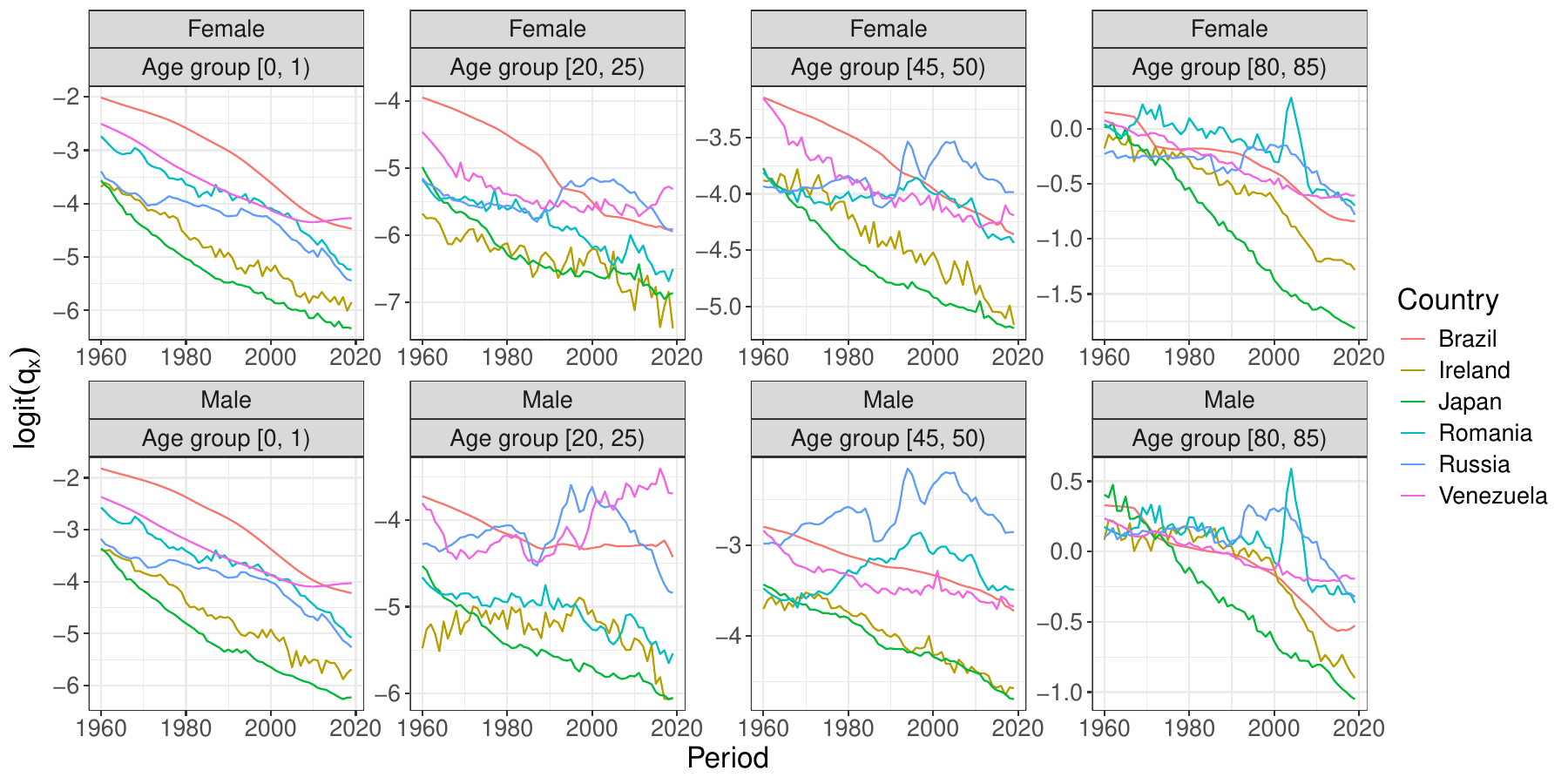}
	\caption{Logit probability of death over time for selected age groups and countries for females and males.}
	\label{fig:data:qx-sample}
\end{figure}

Figure~\ref{fig:data:qx-sample} shows the logit mortality trajectories for selected countries and age groups, providing additional detail on the patterns observed in Figure~\ref{fig:data:tau-wpp}. For females, the mortality trend is negative for the $[0, 1)$ age group across all selected countries, although levels differ considerably, with Brazil and Venezuela exhibiting higher newborn mortality than the other countries. For the young-adult age group $[20, 25)$, we observe differences in levels across countries and a period of mortality increase in Russia, which explains its higher $\tau$ values in Figure~\ref{fig:data:tau-wpp}. For the adult age group $[45, 50)$, Russia and Romania exhibit distinct and increasing mortality trajectories compared with the other countries. Moreover, for this age group and with the exception of Brazil and Venezuela, other countries start at similar levels but diverge shortly after, with Romania eventually reaching mortality levels similar to those of Brazil and Venezuela. Finally, for the old age group $[80, 85)$, Russia shows a period of increasing mortality in the 1990s and Romania in the 2000s, but countries differ mainly in terms of levels, with Ireland and Japan achieving low levels at different rates.

For males, we observe a negative trend in newborn mortality in Figure~\ref{fig:data:qx-sample}, although levels vary across the selected countries, with Brazil starting at higher levels than the other countries. For young-adult mortality $[20, 25)$, we observe a more diverse pattern: high levels for Brazil, a recent surge for Venezuela, and increasing mortality for Russia around the 1990s, while Ireland and Japan experience stable or decreasing mortality. For the middle-age group $[45, 50)$, we observe three distinct patterns: a decreasing trend with low levels for Ireland and Japan, a decreasing trend with higher levels for Brazil and Venezuela, and an increasing trend for Russia and Romania (also visible in Figure~\ref{fig:data:tau-wpp}), with the latter two countries exhibiting the highest mortality levels across time for these age groups. For the old age group $[80, 85)$, countries start at similar levels but diverge later, with Ireland and Japan reaching lower levels at different rates.

Overall, we observe diverse patterns of mortality levels and trends across countries, with multiple changes often occurring simultaneously across several age groups. This complexity makes direct analysis of the raw age-specific probabilities of death challenging, 
motivating the need for a more parsimonious summary of how mortality has changed over time, which is the focus of the subsequent sections.

\section{Dimensionality reduction}\label{sec:dim-red}

Mortality probabilities have unit-interval support, high dimensionality across age groups, and strong cross-age correlation structure. To address these characteristics, we adopt the Bayesian time-dependent beta latent variable model (time-BLV) proposed by~\cite{Araujo_2026_blv}. This framework provides a parsimonious representation of age-specific mortality profiles through a low-dimensional latent structure, while explicitly accounting for temporal dependence within countries.

The time-BLV model assumes that $q_{xit} \sim \mathrm{Beta}(\kappa\mu_{xit}, \kappa(1-\mu_{xit}))$ where
\[\mathbb{E}(q_{xit}) = \mu_{xit}, \qquad \text{Var}(q_{xit}) = \frac{\mu_{xit}(1-\mu_{xit})}{\kappa + 1}, \]
with the mean linked through
\[\mathrm{logit}(\mu_{xit}) = \beta_{\mathrm{blv},x} + \boldsymbol{\theta}_{it}^{\top} \boldsymbol{\alpha}_{\mathrm{blv},x}.\]
In the time-BLV model, $\beta_{\mathrm{blv},x}$ represents the baseline age-specific mortality profile, while the $K$-dimensional latent vector $\boldsymbol{\theta}_{it}$ captures country-time deviations from this baseline. The loading vectors $\boldsymbol{\alpha}_{\mathrm{blv},x}$ determine how each latent component influences mortality at age \( x \). This yields a low-rank representation with $K \ll J$, where $J$ denotes the number of age groups. For a single population, the model can be compared with the Lee-Carter model \citep{Lee_Carter_1992}, although the time-BLV model simultaneously models mortality probabilities across multiple populations and allows for more than one latent component.

In the time-BLV model, temporal dependence within each country is introduced via an autoregressive prior:
\[\boldsymbol{\theta}_{it} = \phi_i \boldsymbol{\theta}_{i,t-1}  + \sigma_i \boldsymbol{\epsilon}_{it}, \qquad  \boldsymbol{\epsilon}_{it} \sim \mathcal{N}_K(\mathbf{0}, \mathbf{I}_K),\] which induces smooth trajectories in the latent mortality dynamics. Additionally, as in~\cite{Araujo_2026_blv}, we estimate parameters and latent effects using a Bayesian approach assuming the priors: 
$\phi_i \sim \mathrm{U}(-1,1)$, $\log \sigma_i \sim \mathcal{N}(0,1)$, $\alpha_{\mathrm{blv},xk} \sim \mathcal{N}(0,1)$, $\beta_{\mathrm{blv},x} \sim \mathcal{N}(0,100)$ and $\log \kappa \sim \mathcal{N}(0, 100)$.

The latent dimension $K$ is selected using the Bayesian Information Criterion (BIC) \citep{Schwarz_1978}, computed from an approximation to the marginal log-likelihood as described in \cite{Araujo_2026_blv}, and we retain the model with the smallest BIC. Later, to assess the stability of the clustering results across different values of $K$, we also compute the adjusted Rand index (ARI) \citep{Rand_1971,Hubert_1985} for different numbers of latent effects and numbers of clusters.

\section{Clustering mortality curves and countries}\label{sec:clustering}

Based on the estimated lower-dimensional mortality representation $\hat{\boldsymbol{\theta}}_{it}$ with dimension $K$ for country $i=1,\dots, n$ at time $t=1,\dots, T$, our goal is to identify $G$ clusters for the mortality curves, which we call ``mortality states'' for country $i$ at time $t$, defined as $Z_{it}\in \{1, \dots, G\}$. These mortality states will represent a set of mortality curves with a particular combination of mortality level and shape. 

All mortality states for country $i$ are denoted by the vector $\mathbf{Z}_{i}=(Z_{i1}, \dots, Z_{iT})$, forming a sequence of mortality states, as is common in sequence analysis methods \citep{Billari_2001}. These mortality state sequences $\mathbf{Z}_{i}$ summarise how mortality has been changing over time in each country and are subsequently used to identify $M$ country clusters, grouping countries with similar mortality state evolution. Each country $i$ is assigned to a cluster denoted by $W_i \in \{1, \dots, M\}$.

To quantify temporal patterns in mortality inequality and cycles of divergence and convergence \citep{Vallin_Mesle_2004}, we measure the level of inequality across countries at each period $t$. Given all mortality states across countries for a period $t$, denoted by $\mathbf{Z}_{t}=(Z_{1t}, \dots, Z_{nt})$, we compute the standardised entropy, a metric commonly used in sequence analysis \citep{Gabadinho_2011}. This entropy measure is based on the proportions $p_{tg}$ of countries in each mortality state $g\in\{1,\dots, G\}$ at time $t$ and is defined as:

\[E_{t} = -\frac{1}{\log G}\sum_{g=1}^{G}p_{tg}\log p_{tg}.\]

This multi-stage approach allows us to characterise mortality dynamics at multiple levels: (i) by identifying mortality states $Z_{it}$ that capture distinct mortality profiles at each country-year observation; (ii) by constructing sequences $\mathbf{Z}_i$ and cross-sectional distributions $\mathbf{Z}_t$ to analyse temporal evolution within and across countries; (iii) by grouping countries into clusters $W_1, \ldots, W_n$ 
based on similarities in their mortality state sequences $\mathbf{Z}_i$; and (iv) by measuring temporal inequality through the entropy $E_t$ of the distribution $\mathbf{Z}_{t}$. Together, these components provide a comprehensive description of how mortality has changed over time and which countries have followed similar trajectories.

\subsection{Inferring mortality states}

We infer the mortality states $Z_{it}\in\{1,\dots,G\}$ for each sex by applying k-means clustering to the $K$-dimensional latent effects $\hat{\boldsymbol{\theta}}_{it}$ for all country-year observations $(i,t)$. For a fixed number of clusters $G$, the k-means algorithm assigns observations to clusters by minimising the sum of squared distances to the cluster centroids. Formally, this is achieved by minimising the following objective function:

\[
\sum_{g=1}^{G}\sum_{(i,t)\in C_g} \lVert\hat{\boldsymbol{\theta}}_{it}-\mathbf{c}_g\rVert^2,
\] where $\mathbf{c}_g\in\mathbb{R}^{K}$ is the centroid for cluster $g$, $C_g=\{(i,t): Z_{it}=g\}$ is the set of assigned observations, and $\lVert\cdot\rVert$ denotes the Euclidean distance. 

K-means tends to produce clusters of similar sizes; consequently, the method will capture temporally persistent patterns in the data, while minority patterns present only in a few countries or short periods will tend to be merged into adjacent large clusters.

\subsection{Inferring country clusters}

To cluster countries based on their sequences of mortality states $\mathbf{Z}_{i}$ for $i=1,\dots, n$, we employ a mixture of categorical distributions with cluster-specific time-varying probabilities, referred to as the MixSCat model. The MixSCat allows the probability of occupying each mortality state to vary smoothly over time within each country cluster. Specifically, the model specifies the probability of being in mortality state $g$ at time $t$ given that country $i$ belongs to cluster $h$ as: \[p(Z_{it} = g \mid W_i = h) = \frac{\exp\{\beta_{g} + S_{hg}(t)\}}{\sum_{l=1}^{G}\exp\{\beta_{l} + S_{hl}(t)\}},\] where $\beta_{g}$ is a baseline effect for mortality state $g$, and $S_{hg}(t)$ is a B-spline basis function \citep{Hastie_1990_GAM} of size $L$ (including an intercept) with degree 3, generated by the built-in \texttt{R} package \texttt{splines}. For identifiability, the effect for the last category is zero $\beta_{G} = S_{hG}(t) = 0$ for all clusters, and the spline intercept for the first country cluster is also zero. The spline effect $S_{hg}(t)$ is defined in terms of a fixed B-spline basis function $\mathbf{b}(t)$ and associated spline coefficients $\beta_{hgj}$ that will be estimated, i.e.,
\[S_{hg}(t) = \sum_{j=1}^{L}{b_{j}(t)\beta_{hgj}} = \mathbf{b}(t)^\top\boldsymbol{\beta}_{hg}.\]

In spline modelling, it is common to fix the number of basis functions $L$ and impose smoothness by penalising the spline coefficients. In this work, we adopt a first-order difference prior, specified as $\beta_{hgj} = \beta_{hg,j-1} + \epsilon_{hgj}$ with $\epsilon_{hgj} \sim \mathcal{N}(0, \lambda^{-1})$, which encourages neighbouring coefficients to vary smoothly.

We estimate the model parameters and latent variables within a Bayesian framework, and impose additional regularisation by assuming $\beta_g \sim \mathcal{N}(0,1)$ and fixing $\lambda=1$. This mitigates separability in the clustering model: within a given country cluster, the indicator that a country is in mortality state $g$ at time $t$ can otherwise perfectly predict that country’s mortality state assignment at that same time point, which leads to unstable (potentially unbounded) coefficient estimates.

The prior probability of cluster membership, $p(W_i = m) = \pi_{im}$, is specified in an empirical Bayes fashion, whereby $\pi_{im}$ is estimated from preliminary runs of the model and then treated as fixed in the subsequent analysis. Conditional on these data-driven prior probabilities, full Bayesian inference is carried out by treating the cluster memberships and model parameters as unknown and sampling from their joint posterior distribution. Model estimation is performed via a Markov chain Monte Carlo (MCMC) algorithm, specifically a Gibbs sampler based on Pólya-Gamma augmentation~\citep{Polson_2013}. Further details on the empirical Bayes procedure and the Gibbs sampling scheme are provided in Appendix~\ref{app:estimation-w}.

The MixSCat model offers some advantages for analysing mortality state sequences. First, by grouping countries into country clusters with similar temporal patterns, it provides a clearer visualisation of how mortality states evolve over time within each cluster. Second, the model enables us to identify precisely when major transitions between mortality states occur within each country cluster by examining changes in the modal mortality state probability over time. This temporal specificity is particularly valuable for understanding the timing of mortality transitions and for detecting convergence or divergence patterns across countries.

\subsection{Selecting the number of clusters}\label{subsec:model-selection}

To select the number of mortality states $G$, we compute the average silhouette width (ASW) \citep{Rousseeuw_1987}, a distance-based measure where higher values indicate greater within-cluster cohesion and stronger between-cluster separation. However, the final choice is not determined automatically by the ASW alone. We also consider cluster interpretability by visualising the centroid for each mortality state to verify whether the clusters capture demographically meaningful patterns that align with known mortality dynamics.

For selecting the number of country clusters \( M \), we adopt a sparse finite mixture approach \citep{Rousseau_Mengersen_2011}. Specifically, we assume that the vector of cluster weights $\boldsymbol{\pi} = (\pi_1,\dots,\pi_{M_{\text{max}}})$ follows a symmetric Dirichlet prior, $\boldsymbol{\pi} \sim \text{Dirichlet}(0.1,\dots,0.1)$, and set \( M_{\text{max}} = 15 \), with latent allocations $W_i \mid \boldsymbol{\pi} \sim \text{Categorical}(\boldsymbol{\pi})$, allowing non-active clusters to remain empty. Given the categorical nature of the data and the inherent multimodality induced by the mixture structure, the model is prone to convergence to local modes. To mitigate this issue, we run a large number of short MCMC chains and select the final number of clusters as the most frequently occurring value across these chains. Further details of the algorithm are provided in Appendix~\ref{app:estimation-w}.

\section{Application to the WPP dataset}\label{sec:application}

\subsection{Dimensionality reduction}

We fit the time-BLV model separately for females and males for the WPP data described in Section~\ref{sec:wpp-data}. The BIC values for different $K$ and further details about model fitting are available in Appendix \ref{app:dimred}. For females, the lowest BIC is obtained for $K=4$. For males, the lowest BIC is obtained for $K=5$, followed closely by $K=4$. Overall, for both sexes, after $K > 3$, the resulting mortality states do not differ substantially, as evidenced by similar ARI values. We therefore decide to use $K=4$ dimensions for both females and males, reducing the dimensionality of the data from 18 age groups to 4 latent dimensions, which capture the main features of the mortality curves while accounting for temporal dependence.  

\begin{figure}[H]
	\centering
	\includegraphics[width=0.8\textwidth]{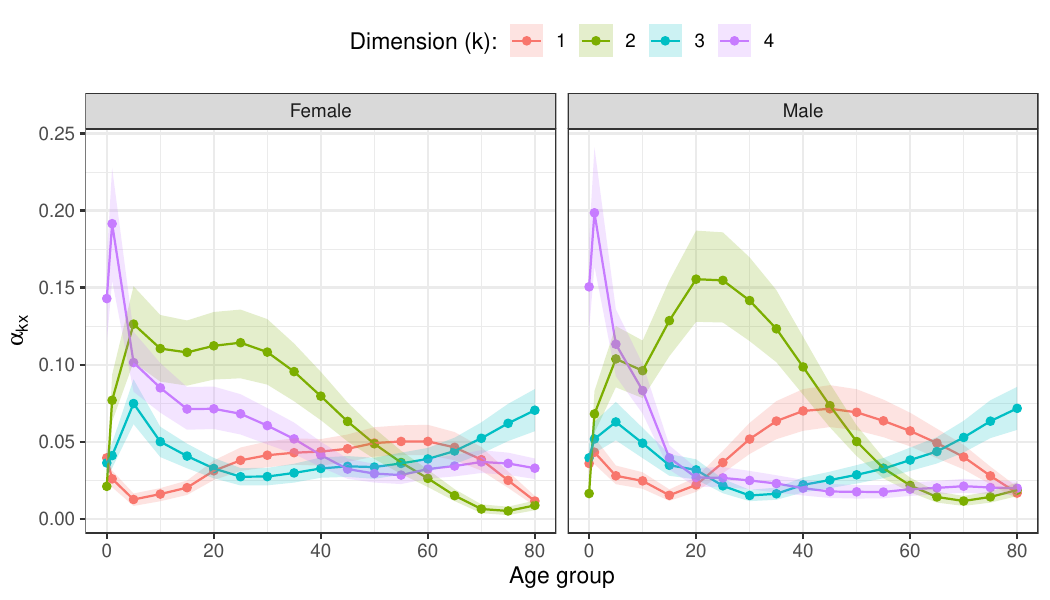}
	\caption{Posterior mean and 95\% high posterior density interval for $\boldsymbol{\alpha}_{\mathrm{blv},x}$ across age groups $x$ and latent dimension $k=1,\dots, 4$ for females and males.}
	\label{fig:dim-red:alpha}
\end{figure}

The posterior mean of the time-BLV coefficients $\boldsymbol{\alpha}_{\mathrm{blv},x}$ for each age group $x$ and dimension $k=1,\dots, 4$ is shown in Figure~\ref{fig:dim-red:alpha}. The coefficients indicate how each latent dimension influences mortality at different ages. For instance, for both females and males, dimension 1 has a stronger effect on late adults and older age groups (40 to 70 years old) and newborns. Dimension 2 affects many young adult age groups (around age 20). Dimension 3 has a greater effect on young and old age groups, while Dimension 4 affects very young age groups.

\subsection{Mortality states}\label{sec:mortality-states-results}

\begin{figure}[H]
	\centering
	\includegraphics[width=0.95\textwidth]{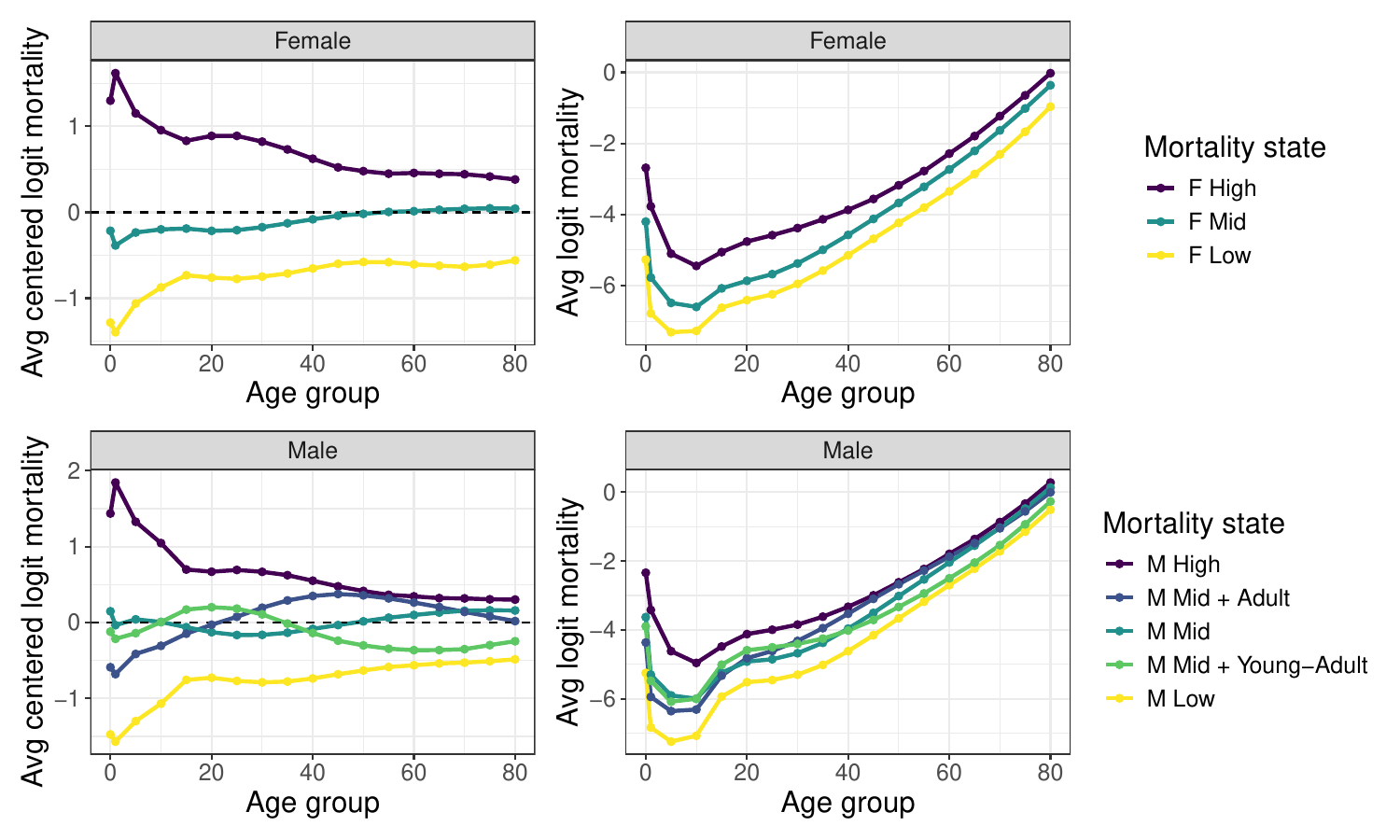}
	\caption{Average (avg) age-centred (on the left) and not centred (on the right) logit mortality curves with 0.025 and 0.975 quantiles for each mortality state for females and males.}
	\label{fig:states-centers}
\end{figure}

We apply $k$-means clustering to the posterior mean of the latent effects for each sex separately, considering $G = 2,\dots,10$, with further details about the fit and ASW available in Appendix~\ref{app::mortality-states}. For females, the highest ASW is for $G=2$ clusters, but this solution reduces all mortality variation to a single binary split around the historical average, providing limited demographic interpretability. We therefore consider $G=3$, the smallest number of clusters that yields demographically meaningful and distinct tiers: a low-mortality state with curves mostly below the historical average (`F Low'); a mid-level mortality state with curves around the average (`F Mid'); and a high-mortality state with curves mostly above the average (`F High'). All three states are shown in Figure~\ref{fig:states-centers}, which displays both centred and uncentred average mortality curves for each mortality state. Below age 40, the three curves are characterised by different levels and shapes, whereas after age 40 the curves look similar but with vertical shifts. For $G > 3$, additional clusters only further subdivide the `F Mid' state based on minor differences in newborn mortality, without introducing new shapes or demographically distinct patterns, so the three-state solution offers a good trade-off between parsimony and interpretability.

For males, the highest ASW is obtained for $G=3$ clusters, followed by $G=2$ and $G=5$. However, for both $G=2$ and $G=3$, the resulting mortality states differ mainly in level, without capturing distinct age-pattern shapes. The intermediate $G=4$ solution reveals a mortality state with a distinct shape, but with a drop in the ASW when compared with $G = 5$. At $G=5$, two mortality states emerge with distinct age profiles that go beyond level differences. These two shape-based mortality states cannot be recovered at $G < 5$, which is the primary justification for considering $G=5$ for males. For $G > 5$, additional clusters fragment the average-mortality group without producing new meaningful shapes. The resulting five states are: `M Low', with mortality curves mostly below the average; `M Mid', with mortality curves around the average; `M High', with mortality curves mostly above the average; `M Mid + Adult', with mortality curves characterised by lower than average mortality at young ages but higher than average mortality in adult age groups; and `M Mid + Young-Adult', with mortality curves around the average but showing elevated mortality for young adults, while mortality at older ages declines more slowly than average (but remains higher than in `M Low'). The five mortality states are shown in Figure~\ref{fig:states-centers}.

To further summarise the level of each mortality state, we present in Table~\ref{tab:e-cluster} the temporary life expectancies between ages 0 and 85, 20 and 85, and 60 and 85 for each mortality state. The temporary life expectancy between age $x$ (with $x=[0,20,60]$) and 85 is the average number of years that a group of persons alive at age $x$ will live between age $x$ and 85, and it is computed using standard demographic methods \citep{arriaga_measuring_1984}.

For females, the temporary life expectancies decrease from `F Low' to `F Mid' and then `F High' across all three age ranges. The temporary life expectancy between birth and age 85 for `F Low' is approximately 22\% higher than for `F High', corresponding to a difference of 14 years. The differences between the three mortality states are smaller at older ages, with the gap between `F Low' and `F High' decreasing from 14 years between ages 0 and 85 to 4 years between ages 60 and 85. `F Mid' consistently has intermediate temporary life expectancies between `F Low' and `F High'.

For males, the gap in temporary life expectancy between birth and age 85 between `M Low' and `M High' is 17 years. `M Mid' and `M Mid + Adult' have similar temporary life expectancies between birth and age 85 (68 and 67 years, respectively), but between ages 20 and 85, temporary life expectancy is approximately 2 years lower for `M Mid + Adult'. The `M Mid + Young-Adult' mortality state has slightly higher temporary life expectancies than `M Mid' and `M Mid + Adult' across all three age ranges, but lower than `M Low'. `M Mid' has temporary life expectancies that are close to the average of `M Low' and `M High', particularly for the older age ranges.

\begin{table}[H]

\centering

\caption{Temporary life expectancy (years) between ages 0 and 85, 20 and 85, and 60 and 85 by mortality state and sex.}

\vspace{0.25cm}

\label{tab:e-cluster}

\begin{tabular}{
l
S[table-format=2.2]
S[table-format=2.2]
S[table-format=2.2]
}

\toprule

& \multicolumn{3}{c}{Age} \\

\cmidrule(lr){2-4}

Mortality state & {0} & {20} & {60} \\

\midrule

\multicolumn{4}{l}{\textit{Females}} \\

F Low  & 78.27 & 58.93 & 20.82 \\
F Mid  & 73.68 & 55.32 & 18.33 \\
F High & 64.26 & 51.10 & 16.42 \\

\addlinespace

\multicolumn{4}{l}{\textit{Males}} \\

M Low               & 74.71 & 55.43 & 18.57 \\
M Mid               & 67.55 & 50.25 & 15.32 \\
M Mid + Young-Adult & 70.21 & 52.57 & 17.69 \\
M Mid + Adult       & 66.87 & 48.33 & 15.07 \\
M High              & 57.31 & 46.22 & 14.31 \\

\bottomrule

\end{tabular}

\end{table}

\begin{figure}[H]
	\centering
	\includegraphics[width=\textwidth]{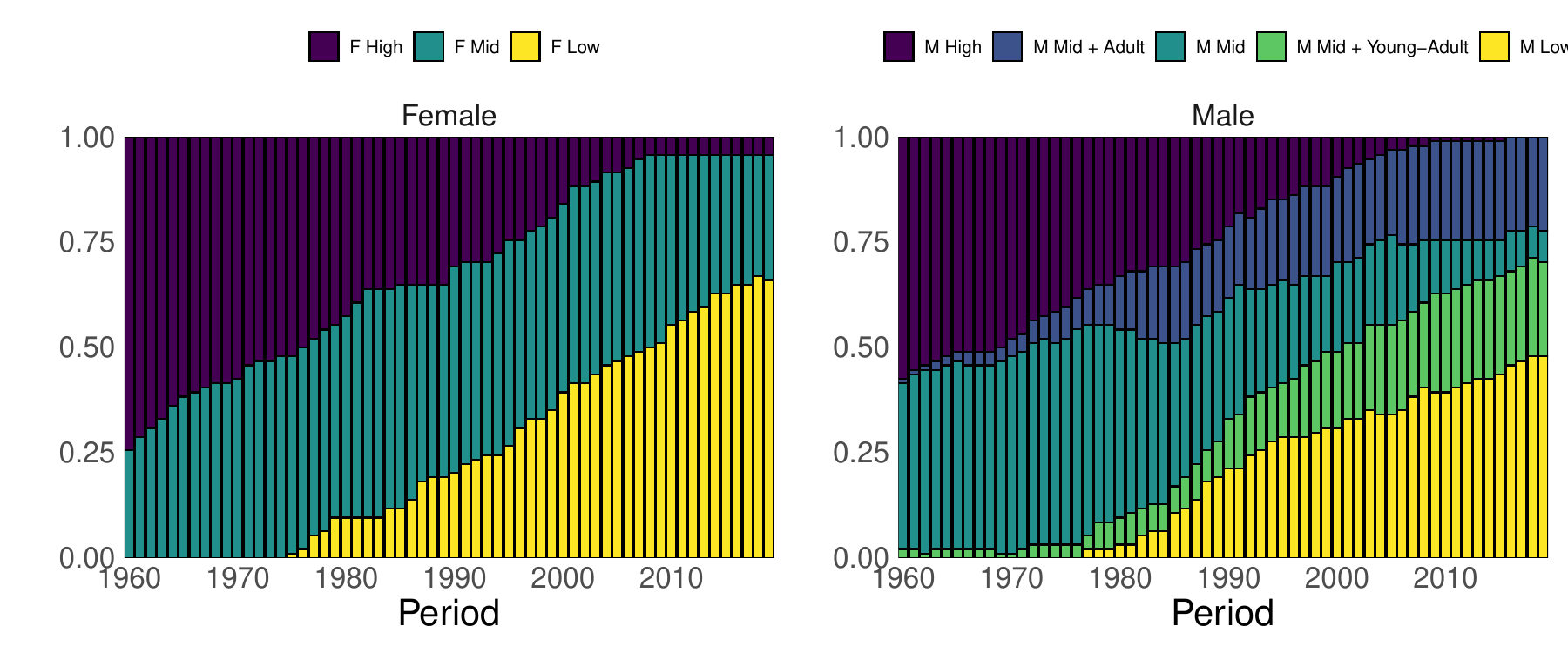}
	\caption{Mortality state proportions over time for females and males.}
	\label{fig:levels-prop-over-time}
\end{figure}

\begin{figure}[H]
	\centering
	\includegraphics[width=0.7\textwidth]{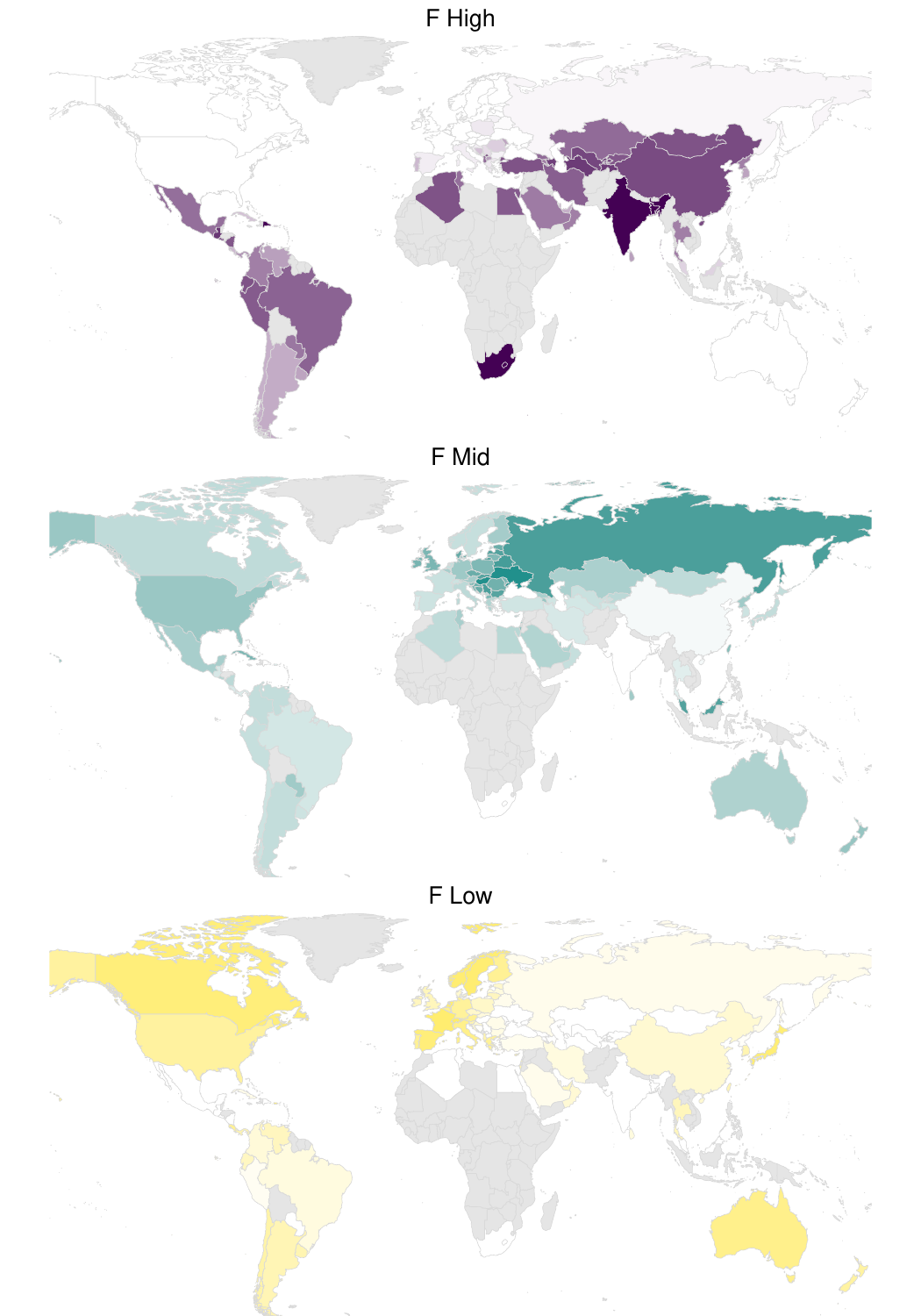}
	\caption{Proportion of time spent in each mortality state per country for females. Colour intensity encodes the proportion, ranging from white (zero) to the darkest shade (one); grey countries were not included in the analysis.}
	\label{fig:fem-state-prop-country}
\end{figure}

\begin{figure}[H]
	\centering
	\includegraphics[width=\textwidth]{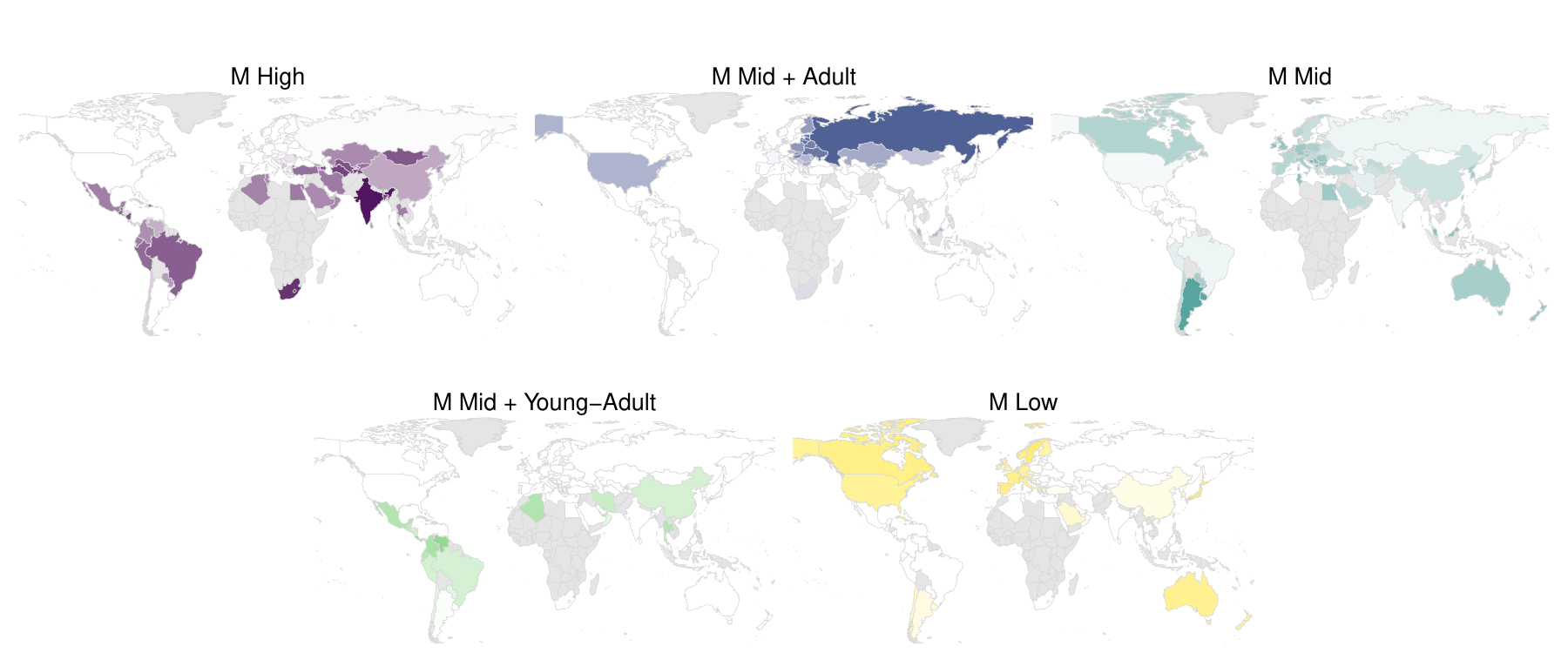}
	\caption{Proportion of time spent in each mortality state per country for males. Colour intensity encodes the proportion, ranging from white (zero) to the darkest shade (one); grey countries were not included in the analysis.}
	\label{fig:mal-state-prop-country}
\end{figure}

While detailed country-level patterns for females and males are analysed in Section~\ref{sec:country-clustes}, the aggregate temporal distribution of mortality states is shown in Figure~\ref{fig:levels-prop-over-time}, and the proportions of time spent in each state across countries are shown in Figures~\ref{fig:fem-state-prop-country} and~\ref{fig:mal-state-prop-country}. 

For females, the period distribution in Figure~\ref{fig:levels-prop-over-time} reveals a progression: the `F High' mortality state dominates at the start of the period, gradually giving way to the `F Mid' mortality state and eventually to the `F Low' mortality state in more recent decades. Males follow a similar trajectory: the `M High' and `M Mid' mortality states are most common in the early period, with countries progressively transitioning through the `M Mid + Adult' and `M Mid + Young-Adult' mortality states before ultimately reaching the `M Low' mortality state.

Regarding the geographical distribution, as seen in Figure~\ref{fig:fem-state-prop-country}, the `F High' mortality state is more prevalent in Latin America, Africa, and Asia. The `F Mid' mortality state appears on every continent, but is the most common mortality state for many Eastern European countries. The `F Low' mortality state is also present in most countries, but is more prevalent in high-income countries. Overall, we observe geographical inequality, with many countries exhibiting a mixture of different mortality states.

For males, as seen in Figure~\ref{fig:mal-state-prop-country}, the `M High' mortality state is more prevalent in Latin America, Africa, and Asia. The `M Mid + Adult' is strongly related to Eastern Europe and Central Asia. The `M Mid' mortality state appears in many parts of the world and is particularly prevalent in Argentina and Uruguay. The `M Mid + Young-Adult' is prevalent in Latin America, while the `M Low' mortality state is mostly present in high-income countries.

\begin{figure}[H]
	\centering
	\includegraphics[width=0.7\textwidth]{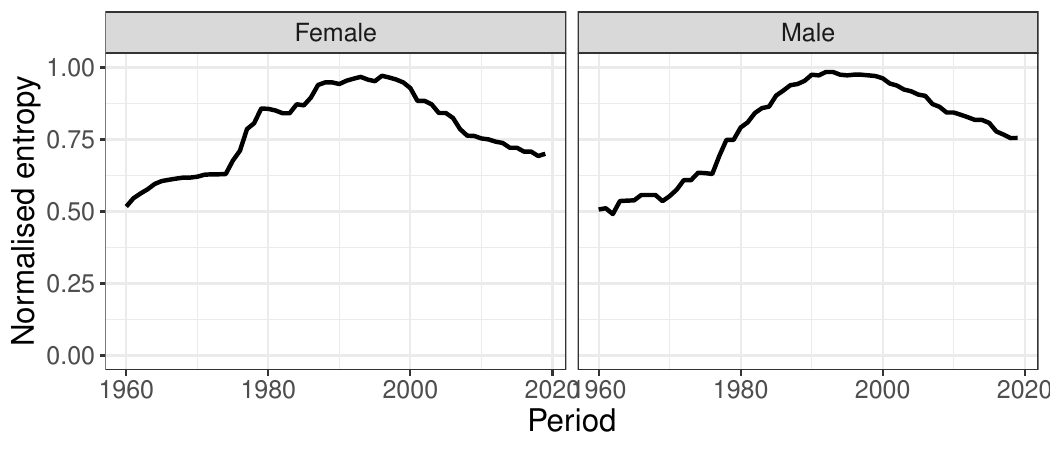}
	\caption{Normalised entropy of the mortality-state distribution over time for females and males under different numbers of mortality states.}
	\label{fig:entropy}
\end{figure}

To quantify mortality inequality across countries over time, we compute the normalised entropy of the mortality-state distribution, as shown in Figure~\ref{fig:entropy}. This measure takes values in the $[0,1]$ interval and equals zero when all countries are in the same mortality state at a given period, and one when mortality states are equally represented in that period. For both sexes, entropy increases from the 1960s to the 1990s, indicating growing dispersion as countries occupy a more diverse set of mortality states. However, after peaking in the 1990s, entropy declines, suggesting recent convergence as countries become more concentrated in fewer states, predominantly lower-mortality states.

\subsection{Country clusters}\label{sec:country-clustes}

We apply the MixSCat model separately to female and male mortality state sequences. Based on the model selection described in Section~\ref{subsec:model-selection}, we use $M=12$ clusters for females and $M=14$ for males; further details are available in Appendix~\ref{app::country-clusters}. Final cluster labels were reordered to facilitate comparison across clusters, keeping clusters with similar mortality states adjacent.

\subsubsection{Female country clusters}

\begin{figure}[H]
	\centering
	\includegraphics[width=\textwidth]{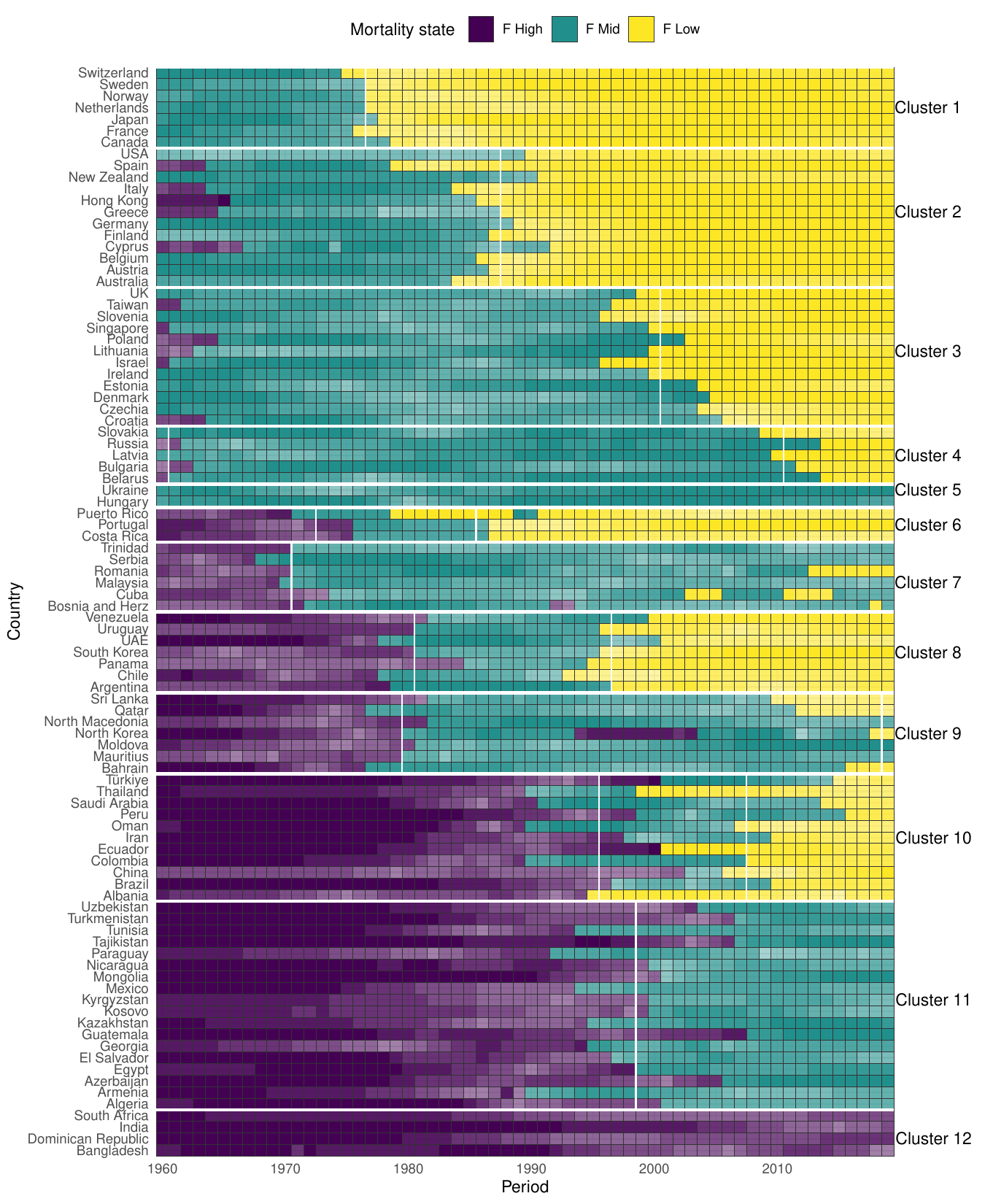}
	\caption[Female mortality states across countries and periods.]{Female mortality states across countries and periods. White horizontal lines separate country clusters, and white vertical lines indicate the time points at which the modal mortality state probability changes within each cluster. Colour intensity reflects the silhouette score of each country-year observation, with darker shading indicating greater separation from the nearest alternative mortality state.}
	\label{fig:seqplot-female}
\end{figure}

Figure~\ref{fig:seqplot-female} displays the mortality state trajectories for each country over time for females, including the resulting country clusters, and the points at which the modal mortality state probability changes within each cluster. The shading of each country-year observation across all periods and countries is based on its silhouette score, and stronger colours indicate that the observation is closer to its assigned mortality state than to the nearest alternative.  

Clusters 1 to 3 contain mostly high-income countries that share a common trajectory, starting in `F Mid' and transitioning quickly to `F Low', although they differ in timing. Cluster 1 contains Canada, France, Japan, the Netherlands, Norway, Sweden and Switzerland and was the earliest to transition, reaching `F Low' by the late 1970s. Cluster 2 contains Germany, Australia, Austria, Spain, the USA, and others and followed roughly 15 years later, around the mid-1990s. Cluster 3 contains Ireland, the UK, Denmark, Czechia, Poland, and many others and achieved this transition only in the early 2000s, representing a gap of over two decades relative to Cluster 1. This split of high-income countries into different female longevity clusters is also consistent with the findings of~\cite{Levantesi_Nigri_2022}.

Clusters 4 and 5 contain Eastern European countries that exhibit more persistent patterns. Cluster 4 contains Russia, Latvia, Belarus, Bulgaria and Slovakia and also started in `F Mid' but remained there longer before eventually transitioning to `F Low' by 2010. Cluster 5 contains Hungary and Ukraine and never left `F Mid' throughout the entire period. The fact that these Eastern European countries are not clustered with Western European and other high-income countries reflects an East-West division that has also been documented in studies such as~\cite{Levantesi_Nigri_2022} and~\cite{Mesle_Vallin_2002}.

Clusters 6 to 9 began in `F High' but transitioned earlier than the remaining low- to middle-income countries, with silhouette scores during `F High' periods suggesting proximity to adjacent mortality states. Cluster 6 contains Costa Rica, Portugal, and Puerto Rico, countries that moved most rapidly across mortality states, reaching `F Mid' by 1970 and then `F Low' around the mid-1980s. Cluster 7 contains Serbia, Malaysia, and others and also transitioned to `F Mid' by 1970, but most remained there. Cluster 8 contains Argentina, Uruguay, Chile, Panama, the UAE and others and followed a similar two-step trajectory, moving to `F Mid' in the late 1970s before reaching `F Low' by the late 1990s. Cluster 9 contains Qatar, Bahrain, North Macedonia, Moldova, and others and transitioned to `F Mid' around the late 1970s, with a few countries reaching `F Low' by 2019. This pattern is in line with studies such as~\cite{Alvarez_Aburto_CanudasRomo_2020}, which show that some Latin American countries, including Costa Rica, Puerto Rico, Cuba, and Uruguay, experienced rapid mortality improvement and approached longevity levels similar to those of high-income countries.

Clusters 10 to 12 contain mostly low- to middle-income countries characterised by prolonged periods in `F High'. Cluster 10 contains Brazil, China, Peru, Iran, Saudi Arabia, and others and demonstrates the most rapid improvement within this group, moving from `F High' to `F Mid' by the late 1990s and ultimately reaching `F Low' by 2009. Cluster 11 contains Mexico, El Salvador, Egypt, Algeria, and many others and achieved a similar transition from `F High' to `F Mid' by the late 1990s but did not progress further. Cluster 12 contains Bangladesh, India and South Africa and remained in `F High' throughout the entire period, though declining silhouette scores in recent years suggest their mortality profiles are approaching those of an adjacent mortality state. The persistently higher mortality levels observed for countries such as Brazil, Colombia, and Mexico relative to high-income countries have also been discussed in~\cite{Alvarez_Aburto_CanudasRomo_2020}.

Taken together, the female country cluster classification reveals substantial global disparity in both mortality states and transition timing. While high-income countries predominantly occupy `F Low' and `F Mid' mortality states throughout the period, middle- and low-income countries spend more time in `F Mid' and `F High'. Transition timing emerges as a key differentiating factor even within income groups, as evidenced by the 20-year gap separating the high-income Clusters 1-3. Moreover, the clusters are not strictly geographically cohesive; Cluster 11, for instance, spans South America, Asia, and Africa, indicating that factors beyond geographic proximity shape female mortality dynamics. 

\subsubsection{Male country clusters}

\begin{figure}[H]
	\centering
	\includegraphics[width=\textwidth]{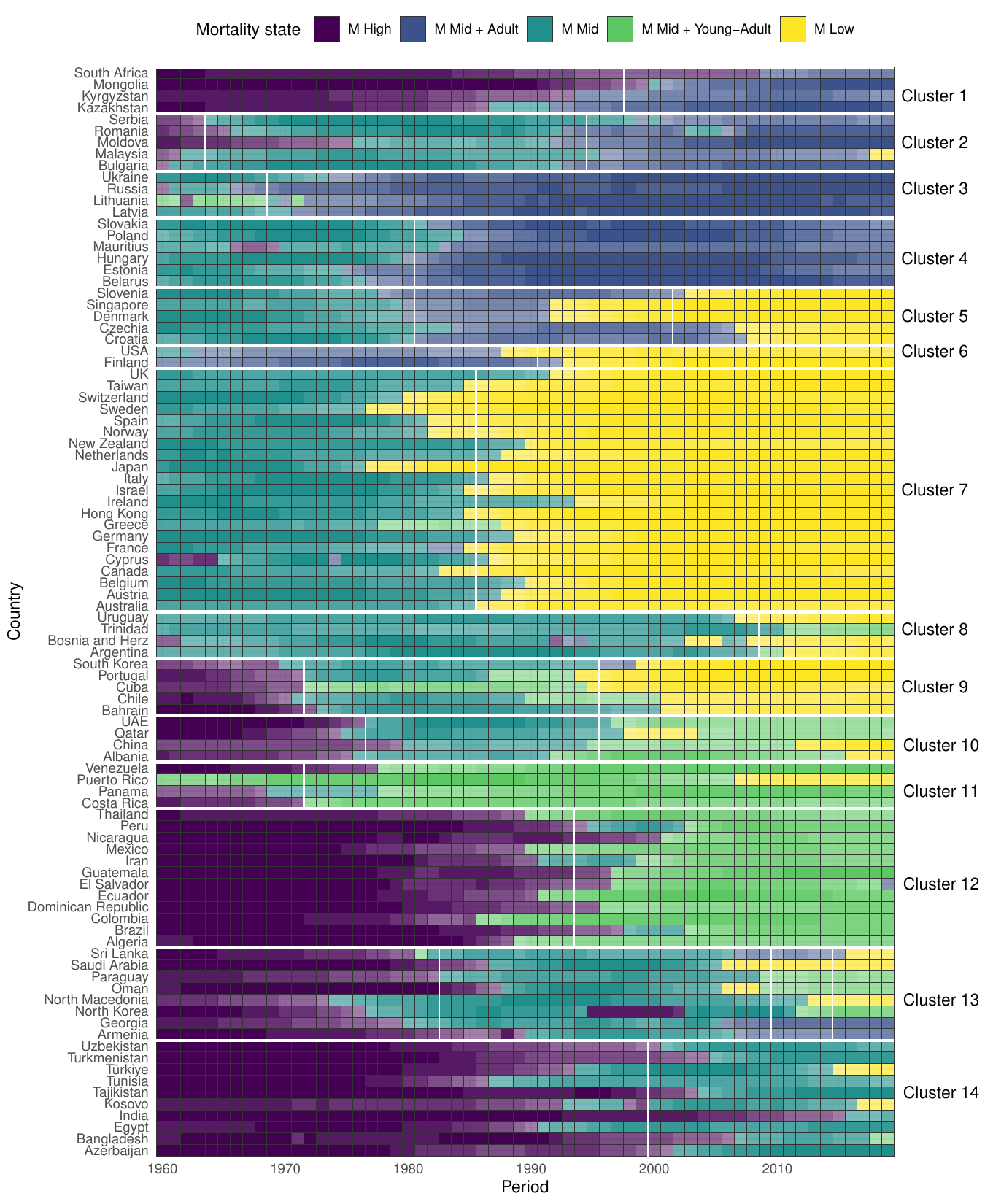}
	\caption[Male mortality states across countries and periods.]{Male mortality states across countries and periods. White horizontal lines separate country clusters, and white vertical lines indicate the time points at which the modal mortality state probability changes within each cluster. Colour intensity reflects the silhouette score of each country-year observation, with darker shading indicating greater separation from the nearest alternative mortality state.}
	\label{fig:seqplot-male}
\end{figure}

As in Figure~\ref{fig:seqplot-female}, Figure~\ref{fig:seqplot-male} shows the male mortality states, country clusters, transition times, and the silhouette scores. Clusters 1 to 4 are characterised by extended periods in the `M Mid + Adult' mortality state. Cluster 1 contains Kazakhstan, Kyrgyzstan, Mongolia and South Africa and started in `M High', moving to `M Mid + Adult' by the late 1990s with low silhouette scores, indicating limited separation from adjacent mortality states. Cluster 2 contains Bulgaria, Moldova, Serbia and Malaysia and moved from `M High' to `M Mid' in the 1960s, then to `M Mid + Adult' in the 1990s. Cluster 3 contains Latvia, Lithuania, Russia and Ukraine and moved from `M Mid' to `M Mid + Adult' by 1970, never leaving this state. Cluster 4 contains Belarus, Estonia, Poland and others and also started in `M Mid' but never left `M Mid + Adult', with more recent periods showing lower silhouette scores. For many countries in these clusters, there is a clear geographical pattern, with Eastern European countries tending to be grouped together, which is a common result in studies clustering male European mortality trajectories~\citep{Mesle_Vallin_2002,Perla_Scognamiglio_2022,Araujo_2025_review}.

Clusters 5 and 6 also contain some periods in the `M Mid + Adult' mortality state, but end in `M Low'. Cluster 5 contains Croatia, Czechia, Denmark, Slovenia and Singapore and started in `M Mid', moved to `M Mid + Adult' by the 1980s, and eventually reached `M Low' by the 2000s. Cluster 6 contains the USA and Finland and started in `M Mid + Adult' with low silhouette scores, indicating weak separation from adjacent mortality states, before moving to `M Low' by 1990. Here, we see some Central European countries grouped separately from the Eastern European countries, a distinction also noted in~\cite{mesle_central_2004} and~\cite{Araujo_2025_review}.

Clusters 7 and 8 share a common trajectory, starting in `M Mid' and ending in `M Low', but differ in timing. Cluster 7 contains mostly high-income countries, including Japan, Canada, France, Italy, and many others, and transitioned to `M Low' by the mid-1980s, making it the largest cluster. Cluster 8 contains Argentina, Uruguay, Trinidad and Tobago, and Bosnia and Herzegovina and followed later, transitioning by 2010. Unlike the female results, where high-income countries were split across different clusters, here they are mostly grouped into a single cluster.

Clusters 9 and 10 started in `M High', then moved to `M Mid' in the 1970s, with divergent outcomes. Cluster 9 contains Chile, Cuba, Portugal and South Korea and ended mostly in the `M Low' mortality state. Cluster 10 contains Albania, China, Qatar, and the UAE and ended in `M Mid + Young-Adult'. In both clusters, silhouette scores are low when in the `M Mid + Young-Adult' mortality state, indicating that these countries are not well-separated from adjacent mortality states during this period.

Clusters 11 to 14 started in `M High' and remained there for an extended period. Cluster 11 contains Costa Rica, Panama, Puerto Rico and Venezuela and transitioned to `M Mid + Young-Adult' around 1970. Cluster 12 contains Brazil, Colombia, Ecuador, Mexico, Peru, as well as Iran, Thailand and others and transitioned to `M Mid + Young-Adult' by the mid-1990s. Cluster 13 contains Armenia, Georgia, Oman, Paraguay and others and transitioned to `M Mid' by the 1980s and then to a mix of `M Mid + Young-Adult' and `M Low'. Cluster 14 contains Bangladesh, Tunisia, India, Egypt and others and transitioned to `M Mid' by 2000. For Latin American countries, this ordering is consistent with~\cite{Revuelta_Hidalgo_2020}, where countries such as Brazil, Colombia, and Mexico historically appear in a worse mortality position than countries such as Costa Rica, Puerto Rico, and Cuba when compared to high-income countries.

Taken together, the male country clusters show substantial diversity in both mortality states and transition periods. Low- and middle-income countries have more `M High' mortality states, while high-income countries have more `M Low' mortality states. Latin American countries are split across clusters based on when they reached `M Mid + Young-Adult'; similarly, Eastern European and some Central Asian countries are split based on when they transitioned to and out of `M Mid + Adult'.

\section{Discussion}\label{sec:conclusion}

This work has introduced a multi-stage clustering framework to analyse human mortality dynamics, inequality, and between-country similarities, which we have illustrated using WPP data. Our framework is based on three steps. First, we applied a dimensionality reduction model to identify smooth time-dependent latent effects. Second, based on the estimated latent effects, we identified discrete mortality states over time that summarise age-specific mortality levels and shapes. Finally, we clustered countries based on their sequences of mortality states, using a smooth, time-varying description of the probability of occupying each mortality state within each country cluster.

Unlike approaches that summarise entire mortality trajectories with a single distance measure, which make it difficult to understand when and why countries differ, our method clusters countries using temporal sequences of mortality states. This approach yields a more interpretable and time-explicit summary of how mortality has evolved within and across countries, capturing both transitions between mortality states and the timing of those changes.

The three mortality states identified for females exhibit distinct levels and shapes at younger ages, with shapes tending to converge at older ages, where differences are primarily observed in mortality levels. For males, on the other hand, 
our analysis uncovered five mortality states characterised by greater variability across all age groups than for females, including two states that we denoted as `M Mid + Adult' and `M Mid + Young-Adult'. The `M Mid + Adult' mortality state, more prevalent in Eastern European and Central Asian countries, 
could be linked to the fall of the Soviet Union, different experiences of the cardiovascular revolution, and social and economic problems~\citep{Mesle_Vallin_2002, Vallin_Mesle_2004}; the `M Mid + Young-Adult' mortality state, mostly associated with Latin American countries, could be linked to violence-related mortality, a well-known problem in this region~\citep{Canudas_Romo_2019,Dvila_Cervantes_2024}. It is worth noting that the lack of `M Mid + Adult' and `M Mid + Young-Adult' mortality states for females does not necessarily mean that female mortality was not affected by these phenomena, and we actually observed some visual effect in Figure~\ref{fig:data:qx-sample}, but it was not large and widespread enough to form a distinct mortality state cluster under k-means.

Despite differences in the number of mortality states employed, both sexes exhibit a similar cross-country inequality pattern, with an initial increase followed by a decline over the study period, a pattern consistent with the divergence/convergence framework \citep{Vallin_Mesle_2004}. Specifically, cross-country inequality in mortality increased until the 1990s, reflecting the coexistence of multiple mortality states and pronounced divergence within both Europe and Latin America, and then began a slow but steady decline.

For the female country clusters, the timing of mortality states played a major role in splitting countries, given the small number of mortality states. For instance, for high-income countries (mostly European), a gap of almost 20 years is observed in the transition to `F Low' when comparing Switzerland with Ireland and the UK, a pattern that aligns with findings from~\cite{Levantesi_Nigri_2022}, who identified similar heterogeneity when clustering female life expectancy trajectories using HMD data. Female country clusters are also not always geographically cohesive, indicating more complex female mortality dynamics.

For males, the country clusters are more geographically cohesive, mostly driven by a more diverse and localised presence of mortality states (`M Mid + Young-Adult' for Latin America, and `M Mid + Adult' for Eastern Europe). The East-West European division was clearly captured \citep{Mesle_Vallin_2002}, and we see that the characteristic higher-than-average adult mortality was also present in countries such as Kyrgyzstan, Kazakhstan, Moldova and Malaysia. Additionally, we observed a diverse classification in Latin America, with countries from the region splitting across four different country clusters varying in timing and mortality states.

Regarding limitations, the mortality states and country clusters identified are specific to the set of countries and period analysed, and do not necessarily generalise to other sets of countries and periods, where the results would likely differ. There is no single ``correct'' choice of countries and periods, as this mainly depends on data availability and research questions. Another limitation concerns the data used in our study. To provide a global analysis, we employed data from the World Population Prospects, restricting the analysis to countries classified as ``Empirical'' and excluding those whose mortality patterns are model-based because of insufficient data. Nevertheless, data for some ``Empirical'' countries are subject to adjustments, which may influence the resulting curve classifications. However, because the UNPD considers the available empirical data for these countries to be sufficiently extensive and reliable, with any adjustments playing only a minor role, and because our analysis excludes the oldest age groups, we believe that the impact of this limitation is likely to be limited.    

Our methodology also has limitations. K-means tends to find clusters of similar size and, combined with our long study period, therefore mostly identifies groups that are sufficiently large to form clusters. Therefore, we may overlook subtle patterns of potential interest to some researchers, such as more recent mortality trends that are unlikely to emerge under our methodology and would require a smaller set of periods or countries, or a different approach. Additionally, our methodology for the country clusters does not account for the uncertainty in the classification of the mortality states, ignoring the fact that some mortality curves most likely lie between a number of mortality states. Developing a unified model that finds mortality states and country clusters in a single probabilistic model by, for instance, assuming a Gaussian mixture model for the mortality states \citep{Raftery_2007} could address this limitation.

Future research could address some limitations by unifying all steps into a joint model that captures dependencies between mortality state allocation, country clusters, and dimensionality reduction. For instance, a hierarchical framework based on a mixture of common factor analysers \citep{Baek_2010} could enable joint estimation of latent effects, mortality states, and country clusters. Beyond methodological improvements, including more countries in the analysis and testing the robustness of the results would be valuable for providing a more comprehensive understanding of global mortality dynamics. Additionally, applying the framework to subnational data could reveal within-country heterogeneity in mortality dynamics, which is increasingly relevant given the growing availability of such data. Extending the analysis to include cause-of-death data could provide new evidence on health transition patterns \citep{frenk_elements_1991,Vallin_Mesle_2004}.

In summary, we present a data-driven approach to summarise mortality dynamics using a multi-stage clustering framework that captures diversity in level, shape, and timing. Promising directions for future work include methodological integration, robustness checks, and extending coverage to additional countries or alternative datasets, including subnational data.

\bibliographystyle{apalike}
\bibliography{references}  

\begin{appendices}

\section{Estimation of country clusters}\label{app:estimation-w}

To express the MixSCat model in matrix form, let $\mathbf{B}$ denote the $T \times L$ matrix of B-spline basis functions evaluated at times $t = 1, \dots, T$, where each row corresponds to the vector $\mathbf{b}(t) \in \mathbb{R}^{L}$ of basis functions at time $t$. For each cluster $h = 1, \dots, M$ and mortality state $g = 1, \dots, G$, let $\boldsymbol{\beta}_{hg}$ denote the vector of spline coefficients. The baseline mortality state effects are defined separately as scalars $\beta_{g}$ for $g = 1, \dots, G-1$. For identifiability, the coefficients corresponding to the reference category $g = G$ are fixed to zero, i.e., $\beta_{G} = 0$ and $\boldsymbol{\beta}_{hG} = \mathbf{0}$ for all $h$. In addition, for the first cluster ($h = 1$), the coefficient vectors $\boldsymbol{\beta}_{1g}$ exclude the first element, so that the intercept is not redundantly parameterised. We collect all coefficients into the $LM \times G$ matrix $\boldsymbol{\beta}$ by stacking the intercepts in the first row and the spline coefficients by cluster:
\[
\boldsymbol{\beta} =
\begin{bmatrix}
\beta_{1} & \beta_{2} & \cdots & \beta_{G-1} & 0 \\
\boldsymbol{\beta}_{11} & \boldsymbol{\beta}_{12} & \cdots & \boldsymbol{\beta}_{1,G-1} & \mathbf{0} \\
\boldsymbol{\beta}_{21} & \boldsymbol{\beta}_{22} & \cdots & \boldsymbol{\beta}_{2,G-1} & \mathbf{0} \\
\vdots & \vdots & \cdots & \vdots & \vdots \\
\boldsymbol{\beta}_{M1} & \boldsymbol{\beta}_{M2} & \cdots & \boldsymbol{\beta}_{M,G-1} & \mathbf{0}
\end{bmatrix}.
\]
The first row contains the intercept terms for each non-reference category, while each subsequent row corresponds to the cluster-specific spline coefficients. The final column is set to zero to enforce the baseline-category constraint.

To avoid overfitting, the spline coefficients are penalised with a first-order difference, where $\beta_{hgj} = \beta_{hg,j-1} + \epsilon_{hgj}$ and $\epsilon_{hgj} \sim \mathcal{N}(0, \lambda^{-1})$. In a Bayesian formulation, this penalisation is equivalent to an improper prior distribution, with a precision matrix given by $\mathbf{P} = \mathbf{D}^\top\mathbf{D}$, where $\mathbf{D}$ is an $(L-1) \times L$ matrix representing the first-order difference:

\[\mathbf{D} = \begin{bmatrix}
-1 & 1 & 0 & 0 & \cdots &  0 & 0 \\
0  &-1 & 1 & 0 & \cdots &  0 & 0\\
0  &0 & -1 & 1 & \cdots &  0 & 0\\
\vdots & \vdots & \vdots & \vdots & \ddots &  \vdots & \vdots \\
0 &  0 & 0 &  0 & \cdots & -1 & 1\\
\end{bmatrix}.\]
To avoid the improper prior, we use a mixed-model formulation, which provides a proper prior for the spline penalisation term through an orthogonalised basis matrix \citep{Wood_2017_GAM}. We start by applying the QR decomposition to $\mathbf{B} = \mathbf{QR}$, which yields a new penalty matrix $\mathbf{P}^{\star} = \mathbf{R}^{-\top}\mathbf{P}\mathbf{R}^{-1}$. We then apply eigendecomposition to $\mathbf{P}^{\star} = \mathbf{U}\boldsymbol{\Lambda}\mathbf{U}^{\top}$, where $\mathbf{U} = [\mathbf{U}_{0} \mid \mathbf{U}_{+}]$, with $\mathbf{U}_{0}$ being the eigenvector associated with the single zero eigenvalue, and $\mathbf{U}_{+}$ the eigenvectors associated with the positive eigenvalues $\boldsymbol{\Lambda}_{+}$, in increasing order.

Two new matrices are computed: the fixed part $\mathbf{C} = \mathbf{Q} \mathbf{U}_{0}$, which is replaced by a column vector with a single entry of $1$; and the random coefficient data part $\mathbf{V} = \mathbf{Q}\mathbf{U}_{+}$, which has the same scale and is standardised to have variance $1$, incorporating the scale into $\lambda$. The new basis function matrix is then defined as $\mathbf{B}^\star = [ \mathbf{1} | \mathbf{V}]$.

Finally, the design matrix for the model is defined as the Kronecker product $\mathbf{X} = \mathbf{W} \otimes \mathbf{B}^\star$, where $\mathbf{W}$ is a matrix in which each row corresponds to the cluster vector $\mathbf{w}_i$, which takes value 1 in position $h$ if $W_i = h$ and 0 otherwise; the first column of $\mathbf{X}$ is then set to 1, and each row of this matrix is denoted by $\mathbf{x}_{it}$.

\subsection{Gibbs sampler for model parameters}\label{est:gibbs}

We use Bayesian inference \citep{Gelman_2013} to estimate the parameters and the country clusters $W_i$. This approach is facilitated by working with the conditional log-likelihood, as it obviates the need to integrate out the latent variables. Furthermore, the penalisation can be interpreted as a prior distribution. 

To avoid separability problems in categorical data \citep{Mansournia_2017}, where the spline or a group can be a perfect predictor of the category, which can lead to infinite variance, we set the baseline effect as $\beta_g\sim \mathcal{N}(0, 1)$ and $\lambda = 1$. The coefficient matrix $\boldsymbol{\beta}$ has prior precision matrix $\mathbf{S} = \mathbf{I}_{M} \otimes \text{diag}\{(1, 1, \text{diag}\{\mathbf{\Lambda}_{+}\})\}$, where the first element is the precision for the baseline $\beta_{g}$, the second for the spline intercept, and the following for the basis functions, stacked across the $M$ groups.

A Gibbs sampler is implemented based on the Pólya-Gamma augmentation \citep{Polson_2013}. Given that the Pólya-Gamma is designed for likelihoods with a binomial form, we need to reparameterise the multinomial distribution, as done in \cite{Held_Homes_2006}. Defining $\boldsymbol{\beta}_{g}$ as the $g$th column of the matrix $\boldsymbol{\beta}$, and $\mathbf{x}_{it}$ as the rows of $\mathbf{X}$ for country $i$ at time $t$, we have:

\[p(\boldsymbol{\beta}_{g} \mid \boldsymbol{\beta}_{-g}, \mathbf{X}, \mathbf{Z}, \mathbf{W})\:\propto\: \prod_{i=1}^{n}\prod_{t=1}^{T}[\eta_{itg}]^{I(Z_{it}=g)}[1-\eta_{itg}]^{I(Z_{it}\neq g)},\] where \[\eta_{itg} = \frac{\exp\{\mathbf{x}_{it}\boldsymbol{\beta}_{g} - U_{itg}\}}{1 + \exp\{\mathbf{x}_{it}\boldsymbol{\beta}_{g} - U_{itg}\}}, \:\text{ and }U_{itg} = \log \sum_{k\neq g}\exp\{\mathbf{x}_{it}\boldsymbol{\beta}_{k}\}.\]

Let $\mathbf{Z}_g$ denote the vector with entries $\mathbb{I}(Z_{it}=g) - 0.5$ stacked over $(i,t)$, let $\mathbf{U}_g$ collect the corresponding $U_{itg}$ values, and let $\boldsymbol{\omega}_g$ be the vector of Pólya--Gamma latent variables \citep{Polson_2013}, and $\Omega_{g} = \text{diag}\{\boldsymbol{\omega}_g\}$. The conditional posteriors are then given by \[\boldsymbol{\omega}_g \,\mid\, \mathbf{X}, \boldsymbol{\beta}, \mathbf{Z}, \mathbf{W} 
\sim \text{PG}\bigl(1, \, \mathbf{X}\boldsymbol{\beta}_{g} - \mathbf{U}_g\bigr),\] \[\boldsymbol{\beta}_{g} \,\mid\, \boldsymbol{\omega}_{g}, \mathbf{X}, \boldsymbol{\beta}_{-g}, 
\sim \mathcal{N}_{LM}(\mathbf{m}_{\omega_g}, \mathbf{V}_{\omega_g}).\] where $\text{PG}$ denotes the Pólya--Gamma distribution, $\boldsymbol{\beta}_{-g}$ is the matrix of coefficients without mortality state $g$, and \[\mathbf{V}_{\omega_g} = (\mathbf{X}^{\top}\Omega_{g}\mathbf{X} + \mathbf{S})^{-1},\] \[\mathbf{m}_{\omega_g} = \mathbf{V}_{\omega_g} \, \mathbf{X}^{\top}\bigl(\mathbf{Z}_g + \Omega_{g}(\mathbf{X}\boldsymbol{\beta}_g-\mathbf{U}_g)\bigr).\]

The full conditionals for the cluster allocation are given by \[p(W_{i} = h \mid\mathbf{Z}_i, \boldsymbol{\beta}, \boldsymbol{\pi}) \:\propto\: \pi_{h}\prod_{t=1}^{T}p(Z_{it} \mid\boldsymbol{\beta}, W_i = h).\]

Let $\boldsymbol{\pi} = (\pi_{1}, \dots, \pi_{M})$ denote the vector of prior cluster probabilities, which can be fixed or random. If the prior probability follows a symmetric Dirichlet distribution $\boldsymbol{\pi} \sim \text{Dirichlet}(a, \dots, a)$, then the full conditional is \[\boldsymbol{\pi} \mid \mathbf{W} \sim \text{Dirichlet}(n_{1} + a, \dots, n_{M} + a),\] where $n_{h}$ is the number of countries allocated to cluster $h$.

\subsection{Model selection and initialisation}

For model selection, we adopt a sparse Dirichlet prior $\text{Dirichlet}(0.1, \dots, 0.1)$ on the cluster weight vector $\boldsymbol{\pi}$, and set an upper bound of $M = 15$ clusters. Given the categorical nature of the data and the multimodality induced by the mixture structure, the model is prone to becoming trapped in local modes. To mitigate this issue, we run $n_{\text{chains}} = 200$ short MCMC chains with $n_{\text{init}} = 50$ iterations from randomly initialised cluster allocations and define the final number of clusters, denoted by $M_{+}$, as the most frequently occurring number of non-empty clusters across the chains.

We then refit the model using $M_{+}$ clusters. To further reduce sensitivity to local modes, we incorporate information from previous runs in an empirical Bayes fashion. Specifically, we extract the final cluster allocations $W_i^{(l)}$ from each chain $l$ and select as a reference partition $\widehat{W}_i$ the one that maximises the posterior expected Rand index (PEAR) \citep{Fritsch_Ickstadt_2009}, computed using the co-clustering matrix across chains. The samples $W_i^{(l)}$ are subsequently relabelled to match $\widehat{W}_i$ using the Equivalence Classes Representatives (ECR) method from the \texttt{label.switching} package \citep{Papastamoulis_2016}. Based on the relabelled samples, we estimate the empirical cluster membership probabilities $\hat{\pi}_{ih}$, which are then used to define the prior distribution $W_i \sim \text{Categorical}(\hat{\pi}_{i1}, \dots, \hat{\pi}_{iM_{+}})$ and the initial values $W_i^{(0)}$, taken as the cluster assignment that maximises the empirical probabilities $(\hat{\pi}_{i1}, \dots, \hat{\pi}_{iM_{+}})$. This choice allows countries that were frequently clustered together in the initial runs to retain a non-negligible probability of belonging to the same cluster, thereby reducing the strong separability behaviour induced by the spline specification and the categorical nature of the data.

\section{Dimensionality reduction}\label{app:dimred}

We set $K=1,\dots,10$, using 15,000 iterations with 5,000 warm-up iterations, a thinning factor of 25, and 3 chains, resulting in a total posterior sample size of 1,200. Convergence assessments such as the $\hat{R}$ and the effective sample size (ESS) \citep{Gelman_2013} of the chains are available in Appendix \ref{app:dimred}. We estimate the parameters by computing the mean of the posterior samples for each parameter.

\begin{figure}[H]
	\centering
	\includegraphics[width=0.9\textwidth]{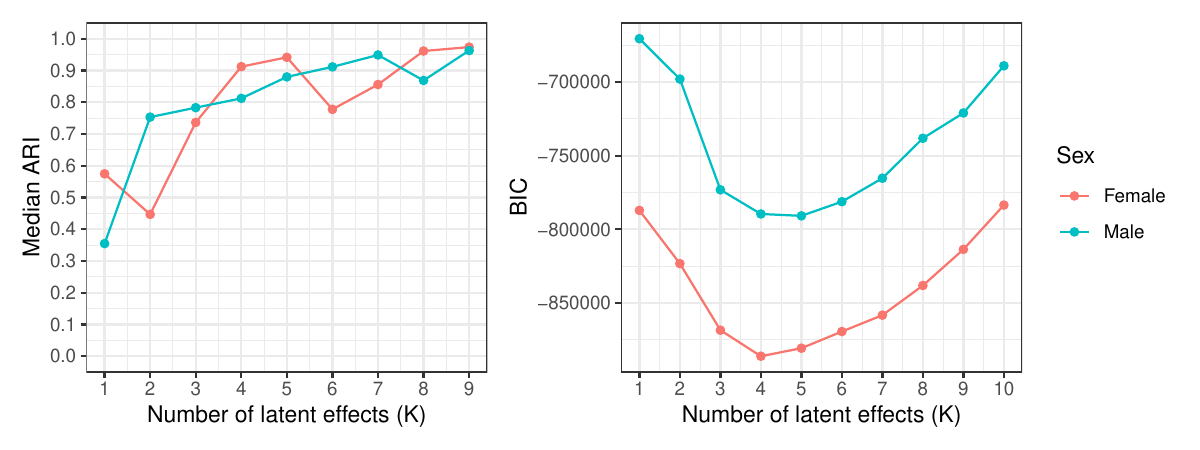}
	\caption{On the left, the median ARI across $G=1,\dots,10$ clusters using k-means for $K$ and $K+1$ latent effects for each sex; on the right, the BIC for each sex across $K=1,\dots, 9$.}
	\label{fig:dim-red:rand-index}
\end{figure}

Figure \ref{fig:dim-red:rand-index} shows the BIC and the average ARI between $K$ and $K+1$ across $G=1,\dots, 10$. Table \ref{r-tblv} shows the $\hat{R}$ and Table \ref{ess-tblv} the ffective sample size for time-dependent beta latent variable model fit for female and males. 

\begin{longtable}{@{\extracolsep{\fill}} c l r r r r}

\caption{Maximum $\hat{R}$ convergence diagnostics by parameter group, sex, and number of latent dimensions $K$ for the time-depenedent beta latent variable model parameters.}\label{r-tblv} \\

\toprule

$K$ & Sex & $\beta$ & $\alpha$ & $\theta$ & $\kappa$ \\

\midrule

\endfirsthead

\toprule

$K$ & Sex & $\beta$ & $\alpha$ & $\theta$ & $\kappa$ \\

\midrule

\endhead

\bottomrule

\endfoot

1  & Female & 1.113 & 1.062 & 1.269 & 1.051 \\
   & Male   & 1.001 & 1.003 & 1.009 & 1.003 \\

2  & Female & 1.159 & 1.093 & 1.271 & 1.054 \\
   & Male   & 1.010 & 1.003 & 1.018 & 1.001 \\

3  & Female & 1.010 & 1.032 & 1.038 & 1.003 \\
   & Male   & 1.017 & 1.055 & 1.068 & 1.011 \\

4  & Female & 1.036 & 1.014 & 1.025 & 0.999 \\
   & Male   & 1.020 & 1.025 & 1.027 & 1.000 \\

5  & Female & 1.012 & 1.138 & 1.129 & 1.001 \\
   & Male   & 1.015 & 1.105 & 1.100 & 0.999 \\

6  & Female & 1.023 & 1.096 & 1.100 & 1.001 \\
   & Male   & 1.017 & 1.035 & 1.038 & 1.000 \\

7  & Female & 1.015 & 1.065 & 1.065 & 0.999 \\
   & Male   & 1.008 & 1.059 & 1.058 & 1.003 \\

8  & Female & 1.040 & 1.122 & 1.125 & 1.000 \\
   & Male   & 1.055 & 1.691 & 1.705 & 1.656 \\

9  & Female & 1.017 & 1.074 & 1.084 & 1.000 \\
   & Male   & 1.010 & 1.130 & 1.132 & 1.000 \\

10 & Female & 1.014 & 1.114 & 1.111 & 1.000 \\
   & Male   & 1.050 & 1.039 & 1.051 & 1.003 \\

\end{longtable}

\begin{longtable}{@{\extracolsep{\fill}} c l r r r r}

\caption{Minimum effective sample size (ESS) by parameter group, sex, and number of latent dimensions $K$ for the time-depenedent beta latent variable model parameters.}\label{ess-tblv} \\

\toprule

$K$ & Sex & $\beta$ & $\alpha$ & $\theta$ & $\kappa$ \\

\midrule

\endfirsthead

\toprule

$K$ & Sex & $\beta$ & $\alpha$ & $\theta$ & $\kappa$ \\

\midrule

\endhead

\bottomrule

\endfoot

1  & Female &  57.76 &  34.37 &   8.21 &  441.54 \\
   & Male   & 924.73 & 699.43 & 591.07 & 1299.47 \\

2  & Female &  13.46 &  43.93 &   5.84 &   42.47 \\
   & Male   & 668.19 & 428.08 & 435.64 & 1382.19 \\

3  & Female & 289.58 & 102.02 & 107.94 & 1280.58 \\
   & Male   & 267.60 & 158.56 &  46.36 &  708.77 \\

4  & Female & 116.65 &  66.17 &  65.14 & 1210.83 \\
   & Male   & 244.40 & 158.64 & 171.62 & 1203.25 \\

5  & Female & 176.91 &  18.54 &  18.34 & 1240.04 \\
   & Male   & 227.01 &  21.70 &  23.24 & 1289.56 \\

6  & Female & 160.05 &  35.85 &  27.41 & 1121.24 \\
   & Male   & 335.55 &  62.97 &  59.20 & 1382.69 \\

7  & Female & 240.13 &  50.51 &  43.92 & 1219.87 \\
   & Male   & 247.22 &  59.43 &  57.02 & 1213.86 \\

8  & Female & 155.13 &  15.72 &  16.08 & 1008.52 \\
   & Male   &  83.63 &   3.14 &   3.12 &    3.61 \\

9  & Female & 171.07 &  36.41 &  32.99 & 1181.39 \\
   & Male   & 241.83 &  21.12 &  16.88 & 1203.75 \\

10 & Female & 130.72 &  25.77 &  25.92 & 1126.41 \\
   & Male   &  88.67 &  43.58 &  44.63 & 1304.31 \\

\end{longtable}

\section{Mortality states}\label{app::mortality-states}

We fitted our k-means method with 1{,}000 random initialisations and a maximum of 5{,}000 iterations using the posterior mean latent effects for each sex.

To complement the main-text discussion of cluster quality, we report the average silhouette width (ASW) for each value of $G$. The silhouette score evaluates how well each observation fits its assigned cluster compared with the nearest alternative cluster. For observation $r$, let $a_r$ be the average distance to observations in its own cluster, and let $b_r$ be the smallest average distance to observations in another cluster. The silhouette score is then
\[
s_r = \frac{b_r-a_r}{\max(a_r,b_r)},
\]
which takes values in $[-1,1]$. Values close to $1$ indicate well-separated clusters, values near $0$ indicate overlap between clusters, and negative values suggest possible misclassification. The ASW is the mean of $s_r$ across all observations, so larger ASW values indicate better overall clustering structure.

\begin{figure}[H]
	\centering
	\includegraphics[width=0.9\textwidth]{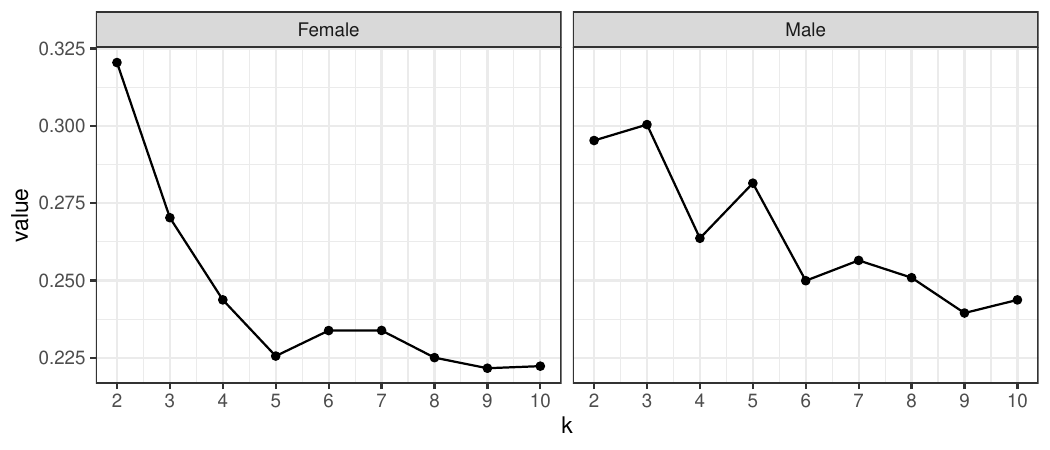}
	\caption{Average Silhouette width (ASW) for $G=2,\dots,10$ mortality states for females and males.}
	\label{fig:mortality-states:metrics}
\end{figure}

\section{Country clusters}\label{app::country-clusters}

We use $L = 10$ spline basis functions, which provide sufficient flexibility to capture temporal patterns while avoiding overfitting and excessive variability in the estimated trajectories. The smoothing parameter is fixed at $\lambda = 1$, which regularises the spline coefficients and ensures stable estimation in the presence of sparse categorical transitions.

\begin{figure}[H]
	\centering
	\includegraphics[width=0.9\textwidth]{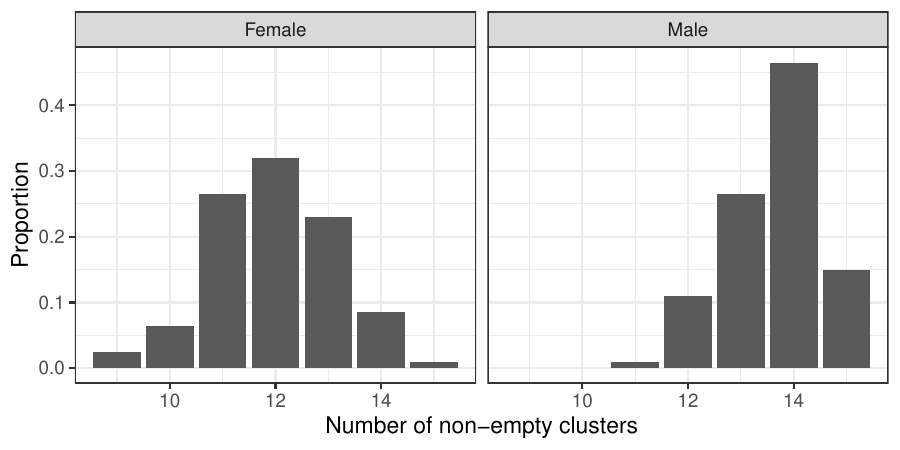}
	\caption{Proportion of non-empty clusters for females and males for the MixSCat model.}
	\label{fig:num-clust}
\end{figure}

For cluster number selection, we employ a sparse Dirichlet prior $\text{Dirichlet}(0.1,\dots,0.1)$ with an upper bound of $M = 15$ clusters. We run 200 short MCMC chains with 50 iterations each, starting from random initial cluster allocations, to explore multiple modes of the posterior distribution. The final number of clusters is chosen as the most frequently occurring number of non-empty clusters across chains, as shown in Figure~\ref{fig:num-clust}.

Conditional on the selected number of clusters, we refit the model using longer MCMC runs with 15{,}000 iterations, a burn-in period of 5{,}000 iterations, thinning every 25 draws, and 3 parallel chains. 

\begin{figure}[H]
	\centering
	\includegraphics[width=0.9\textwidth]{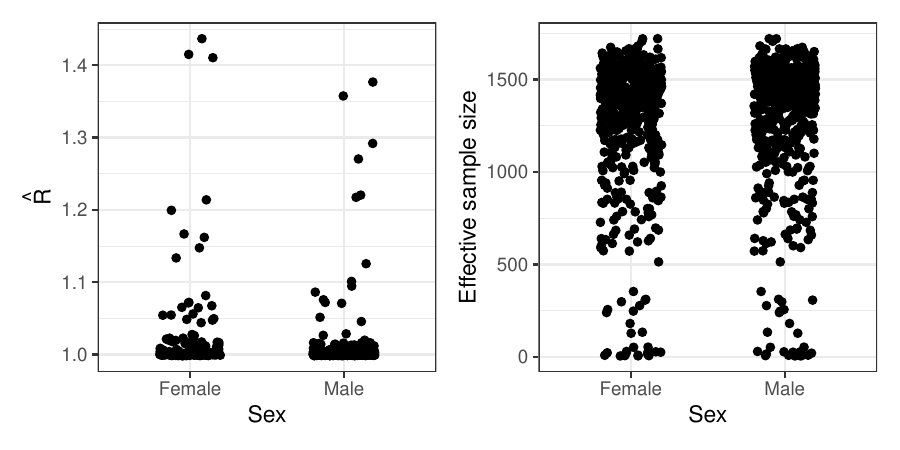}
	\caption{Proportion of non-empty clusters for females and males for the MixSCat model.}
	\label{fig:conv-metrics}
\end{figure}

The $\hat{R}$ and effective sample size are shown in Figure~\ref{fig:conv-metrics}.

\end{appendices}

\end{document}